\documentclass[]{aa} 
\usepackage{graphicx}
\usepackage{color}
\usepackage{txfonts}
\usepackage{hyperref}
\usepackage{siunitx}
\usepackage{natbib}
\usepackage[utf8]{inputenc}
\hypersetup{colorlinks=true, allcolors=blue}
\begin{document} 
\title{HP2 Survey}
\subtitle{V. Ophiuchus: Filament formation in a dispersing cloud complex}
\titlerunning{Feedback driven filament formation}
\authorrunning{Alves et al.}
\author{Jo\~ao  Alves\inst{1}, Marco Lombardi\inst{2}, Charles J.~Lada\inst{3}
}
\mail{joao.alves@univie.ac.at} \institute{%
  University  of Vienna, Department of Astrophysics, T\"urkenschanzstrasse 17, 1180 Vienna, Austria 
  \and University of Milan, Department of Physics, via Celoria 16, I-20133, Milan, Italy 
  \and Harvard-Smithsonian Center for Astrophysics, Mail Stop 72, 60 Garden Street, Cambridge, MA 02138, USA 
  } 
\date{Received: 5 November 2024 ; Accepted: 22 December 2024}

\abstract
 {
 We search for potential ``birthmarks''  left from the formation of filamentary molecular clouds in the Ophiuchus complex. We use high dynamic-range column density and temperature maps derived from \textit{Herschel}, \textit{Planck}, and \textit{2MASS/NICEST} extinction data. We find two distinct types of filaments based on their orientation relative to nearby massive stars: radial (R-type) and tangential (T-type). R-type filaments exhibit decreasing mass profiles away from massive stars, while T-type filaments show flat but structured profiles. We propose a scenario where both filament types originate from the dynamic interplay of compression and stretching forces exerted by a fast outflow emanating from the OB association. The two formation mechanisms leave distinct observable ``birthmarks'' (namely, filament orientation, mass distribution, and star formation location) on each filament type. Our results illustrate a complex phase in molecular cloud evolution with two simultaneous yet contrasting processes: the formation of filaments and stars via the dispersal of residual gas from a previous massive star formation event. Our approach highlights the importance of taking into account the wider context of a star-forming complex, rather than concentrating exclusively on particular subregions.
}  
\maketitle

\section{Introduction}
\label{sec:introduction}
\definecolor{mygrey}{RGB}{180,200,200}

``The region of Rho Ophiuchi is one of the most extraordinary in the sky''; so described E. E. Barnard the great nebula of $\rho$ Oph on plate 13 of his photographic Atlas \citep{barnard1927z}. Barnard's photographs revealed that cloud morphology follows clear patterns, ``the vacant lanes that so frequently run from them [the nebula] for great distances'' \citep{Barnard1907}, or what we call today molecular gas filaments. Filamentary structure in the Interstellar Medium (ISM) was further appreciated and quantified in \cite{Schneider1979} using optical plates, and in every large-scale map of nearby molecular clouds in molecular lines \citep[e.g.,][]{Bally1987,Loren1989a,Goldsmith2008,Soler2018-cc,Soler2021-pu}, dust emission \citep[e.g.,][]{Wood1992,Abergel1994,Johnstone1999}, and dust extinction \citep[e.g.,][]{Cernicharo1985,Lada1994,Alves1998,Lombardi2006,Lombardi2010}. Filamentary structure is also evident in large-scale maps of the diffuse, non-star-forming ISM \citep[e.g.,][]{Heiles1976,Boulanger1985,McClure-Griffiths2006,Soler2022-wh}. More recently, ESA's \textit{Herschel} higher-resolution and sensitivity dust emission maps made it evident that the filamentary nature of clouds extends to sub-pc scales and to a much larger sample of clouds across the Milky Way \citep[e.g.,][]{Miville-Desch2010,Andre2010,Molinari2010}.

\begin{center}
\begin{figure*}[!tbp]

\includegraphics[width=\hsize]{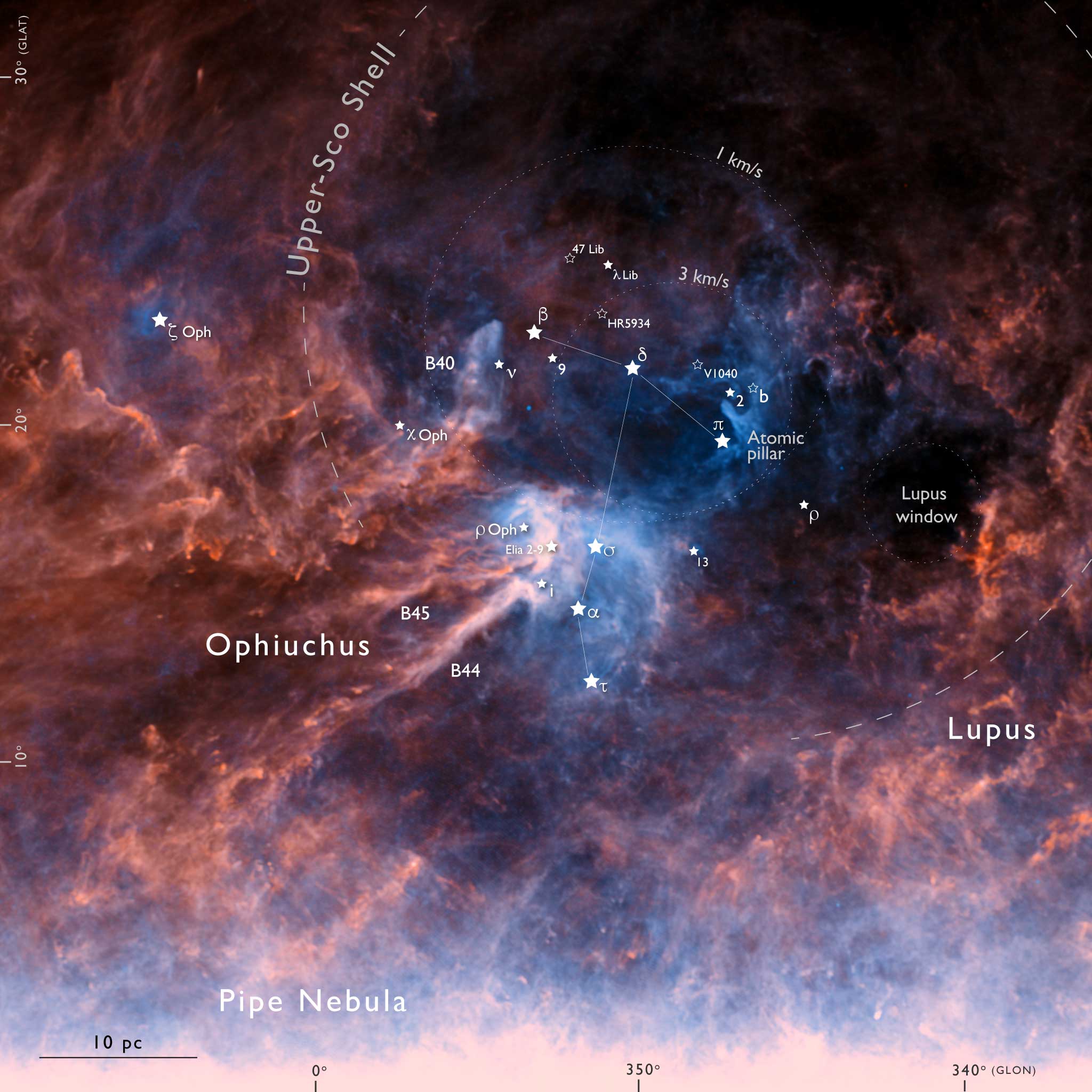}

\caption{Color composite of the Ophiuchus, Pipe Nebula, and Lupus cloud complexes (blue: extrapolated \textit{Planck} 250 $\mu$m, green: \textit{Planck} 350 $\mu$m, red: \textit{Planck} 500 $\mu$m). Star symbols represent the position of massive ionizing stars (B3 or earlier), while symbol size represents relative brightness. Most of these stars have associated H${\alpha}$ extended emission, while only the two fainter objects are not associated with WISE extended emission. Most, if not all stars in this Figure are interacting with the molecular cloud complex and constitute the main source of dust heating (the bluer regions) in the complex.}
\label{fig:planck3color}
\end{figure*}
\end{center}

Filamentary structures are now established as the predominant morphology of the projected gas density field in the ISM, in both atomic and molecular forms \cite[e.g.,][]{Hacar2022-gd,Pineda2023-di}. ISM filaments span a wide range of sizes, from kpc-long filaments \citep[e.g.,][]{Zucker2017-nj} to sub-pc-length \citep[e.g.,][]{Hacar2018-bt}. Although the term ``filament’’ can refer to thread-like structures across various astronomical contexts, we adopt its common ISM definition: projected cloud structures with aspect ratios greater than three. The underlying mechanisms driving the formation of filamentary gas clouds remain the subject of debate and research. Several hypotheses have been proposed. Potential processes responsible for or contributing to filament formation include sheet-like structures seen edge-on, instabilities in such sheets, dynamics of colliding flows, gravitational forces, effects of shocks, the decay of turbulence, feedback, and the magnetic field \citep[e.g.,][]{Nagai1998,Padoan2001,Burkert2004-np,Myers2009,Peretto2012-jd,Heitsch2013-we,Hennebelle2013,Inutsuka2015-io,Tritsis2018,Arzoumanian2018-fj,Bonne2020-we,Jeffreson2020-om,Smith2020-xc,Pillsworth2024-sd}.  

Molecular clouds may retain imprints of their formation history. These ``birthmarks'' could manifest as distinct observational signatures, allowing observers to test formation scenarios.  To investigate this possibility, we need wide-field, high-dynamic range maps of entire cloud complexes. Such maps would enable us to study not only the dense star-forming regions, but also the surrounding diffuse gas that provides crucial context for a cloud's origin. In this paper, we search for such ``birthmarks'' within the filamentary structures of the Ophiuchus molecular cloud complex. Our analysis uses observations from \textit{Herschel}, \textit{Planck}, and \textit{2MASS}, providing a comprehensive view of the cloud across a wide range of densities and temperatures.

\section{The Ophiuchus complex}
The Ophiuchus complex is not an isolated star-forming region, but rather a component of the nearby Sco-Cen OB association, a region that has formed about 13000 stars in the last \SI{20}{Myr}. Most of these stars were formed 15 Myr ago, at the peak of this region star formation rate \citep{Ratzenbock2023-qb}. The OB association also includes the cloud complexes of Lupus, the Pipe Nebula, L134, Corona Australis, and Chameleon. These regions, once thought to be distinct, are now understood to represent the remnants of a larger, single star-forming region \citep{Bouy2015-ce,Ratzenbock2023-qb}. The Ophiuchus complex is embedded within Upper-Sco, a sub-region of Sco-Cen that still harbors approximately 20 massive ionizing stars \cite[e.g.,][]{Miret-Roig2022-hs} and started to form about 10 Myr ago.

A Galactic bubble residing at the boundary between the halo and the Galactic disk in Upper-Sco was identified by \cite{Robitaille2018-om}. This suggests a supernova explosion occurred 1–\SI{3}{Myr} ago, with the resulting shock wave expanding into a pre-existing HI loop created by outflows from the Upper-Sco region.  \cite{Robitaille2018-om} propose that this supernova feedback event triggered the formation of the Ophiuchus and Lupus molecular clouds.  Further supporting this scenario, \cite{Neuhauser2019-ck} found kinematic evidence that the runaway star $\zeta$-Oph and the radio pulsar PSR B1706-16 were ejected from a binary system by a supernova within Upper-Sco approximately \SI{1.8}{Myr} ago.

Recently, \cite{Piecka2024-wg} used ISM optical absorption lines to present evidence for a significant Sco-Cen outflow. The outflow has at least two components: a faster, low-density component traced by Ca II, and a slower, possibly lower-density component traced by Mg II and Fe II. The average radial velocity of this outflow is about -21 km/s and it correlates, partly, with HI emission. A faster flow component, without HI gas, is traceable only in the Ca II line. A flow model suggests an extended distribution of feedback sources within Sco-Cen.

The Ophiuchus region is one of the closest star-forming complexes to Earth \citep[e.g.,][]{deGeus1992,Lombardi2008-qi,Loinard2008,Schlafly2014-tc,Zucker2020-gj}, and, historically, has served as a crucial laboratory to advance our understanding of star formation \citep[e.g.,][]{Struve1948,Blaauw1964,Montmerle1983,Lada1984,Loren1989a,Ward-Thompson1994,Andre1994,Tachihara2000-gf,Tachihara2001-wr}; see \cite{Wilking2008} for a review. This complex exhibits a wealth of filamentary gas structures, mostly cataloged Lynds clouds \citep{Lynds1962} (see Figure~\ref{fig:PHophmap_N}, and can be broadly divided into three subregions:

\begin{enumerate}
    \item Ophiuchus Classic: This region encompasses the L1688 clump (also known as the $\rho$ Oph core), which contains the closest embedded star cluster to Earth, along with the B44 and B45 filaments. These filaments host the actively star-forming regions L1689 \citep[e.g.,][]{Nutter2006-he} and L1709.

    \item Oph North: A filament complex characterized by prominent cores that show limited signs of active star formation \citep{Hatchell12}. 

    \item B40: A less dense and less studied region, also known as the blue horse nebula (illuminated by $\nu$-Sco), including the less known Lynds clouds L1719, L1757, L1782.
\end{enumerate}

Distance estimates to Ophiuchus vary, with \cite{Lombardi2008-qi} reporting \SI{119}{pc} and \cite{Ortiz-Leon2017-em} reporting \SI{137}{pc} and \SI{147}{pc} for L1688 and L1689 (the head of B44), respectively, probably reflecting the 3D structure of the complex. Recent work by \cite{Zucker2019-wr,Zucker2020-gj} found distances ranging from \SI{118}{pc} to \SI{149}{pc} towards the $\rho$~Oph region (L1688) and the Ophiuchus streamers (B44 and B45), with an average distance of \SI{139}{pc}. For simplicity, we adopt an average distance of \SI{140}{pc} throughout this paper. At this distance, \textit{Herschel} dust emission maps achieve resolutions of approximately \SI{4000}{AU} (\SI{0.02}{pc}) at \SI{500}{\micro m} and \SI{2000}{AU} (\SI{0.01}{pc}) at \SI{250}{\micro m}.

In this paper, we analyze the 2D column density map of the Ophiuchus region, created using data from the \textit{Herschel}, \textit{Planck}, and \textit{2MASS} surveys (referred to as HP2), as well as the distribution of massive ionizing stars. Our goal is to uncover signs of the process, or processes, that lead to the formation of filaments in this region. We search for these signs by examining the distribution of mass within the Ophiuchus complex and exploring any potential connections between the arrangement of gas and the presence of massive stars in the region. We find evidence that stellar feedback from massive stars in Upper-Sco played a crucial role in shaping the gas complex in Ophiuchus. This influence extends to the ongoing star formation activity within the association, namely the Lupus cloud complex and the Pipe Nebula.

\section{Data}
\label{sec:data}

The Ophiuchus molecular cloud complex was observed by the all-sky \textit{Planck} observatory and by two \textit{Herschel Space Observatory}  programs, namely, 1) the \textit{Gould Belt Survey} \citep{Andre2010} covering 26.4 sq deg of the complex including L1688, B44 (L1712) and B45 (L1755), and 2) as part of an \textit{Herschel} open time program (PI J. Hatchell), covering 24.2 sq deg towards the Northern regions of the complex (part of B40 and L204).  For \textit{Planck}, the 2013 data products were used \citep{2013arXiv1303.5062P}.  For \textit{Herschel}, we used 160 $\mu$m observations taken in parallel mode using the PACS instrument \citep{Poglitsch2010}, and  250, 350, and 500 $\mu$m observations using the SPIRE instrument \citep{Griffin2010}. Table~\ref{tab:A1} presents a log of the \textit{Herschel} data used in this paper. The data were pre-processed using the \textit{Herschel Interactive Processing Environment} \citep[HIPE][]{2010ASPC..434..139O} version 10.0.2843. A list of the \textit{Herschel} data used in this paper is given in Table~\ref{tab:A1}.

We follow the general procedure introduced in \cite{Lombardi2014} to derive optical depth and temperature maps for the Ophiuchus complex. The procedure can be summarized as follows: 1) We produce an extinction map of the complex using data from the 2MASS-PSC \citep{Skrutskie2006-rn,Lombardi2011-ae} archive and perform a standard reduction of the \textit{Herschel} data with the HIPE; 2) We then convolve the \textit{Herschel} images and re-grid them to match \textit{Planck}'s resolution and data projection; 3) We perform a linear fit between the \textit{Herschel} and \textit{Planck} fluxes at the same passband, confirm the linearity between the two, and we applied the proper calibration offset for each individual \textit{Herschel} passband; 4) We apply the offset to the \textit{Herschel} images and convolve them to the same resolution (normally the resolution of the SPIRE 500 $\mu$m data); 5) We perform an SED fit pixel-by-pixel using a modified black-body as a model, leaving the optical-depth and dust effective temperature as free parameters; the local value of the spectral index $\beta$ is taken from a \textit{Planck}/IRAS fit; 6) Finally, we also build a higher resolution map from the SPIRE 250 $\mu$m band by inferring the optical-depth from the observed flux (and assuming $\beta$ from the \textit{Planck} and $T$ from the 36$^{\prime\prime}$ resolution SED fit).

 % To infer the 3D distribution of the gas in the complex, we used ESA Gaia DR2 astrometric and photometric data \citep{Gaia2018-gb} combined with optical  point  spread  function photometry from the PanSTARRS1 survey \citep{Chambers2016-sq,Magnier2016-fn} and PSF near-infrared photometry from the 2MASS survey \citep{Skrutskie2007-rn}. To infer possible interactions between the most massive stars in the Upper-Sco association and the cloud complex we use WISE satellite infrared data \citep{Wright2010} and WHAM H$\alpha$ data \citep{Haffner2003}.

\section{Results}
\label{sec:results}

In Figure~\ref{fig:planck3color} we present a $\sim30^{\circ}\times30^{\circ}$ square degree color Planck-based composite of the Ophiuchus, Pipe Nebula, and Lupus cloud complexes (blue: extrapolated \textit{Planck} 250 $\mu$m, green: \textit{Planck} 350 $\mu$m, red: \textit{Planck} 500 $\mu$m). The color in this Planck figure corresponds to line-of-sight density weighted temperature (a rough approximation of the true temperature map presented in Figure~\ref{fig:PHophmap_T}) and ranges from about 10 K (orange) to about 30 K (blue).  Star symbols represent the most massive and ionizing stars in the region (spectral type B3 or earlier) and are listed in Table~\ref{tab:ionizing-stars}. Filled star symbols represent stars with extended emission in WISE images, suggesting a close vicinity and interaction with the complex, while open symbols represent stars without recognizable extended WISE emission (see discussion in~\ref{sec:ioniz-stars}). 

\begin{figure}
  \centering
  \includegraphics[width=\hsize]{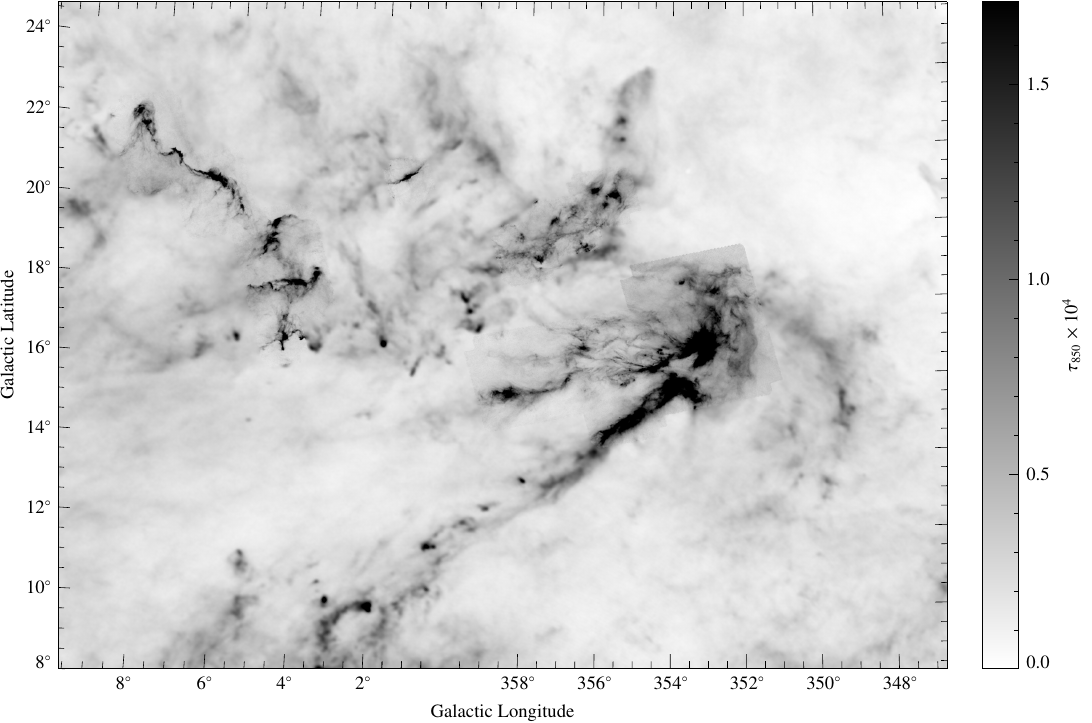}%
  \hspace{-\hsize}%
  \includegraphics[width=\hsize]{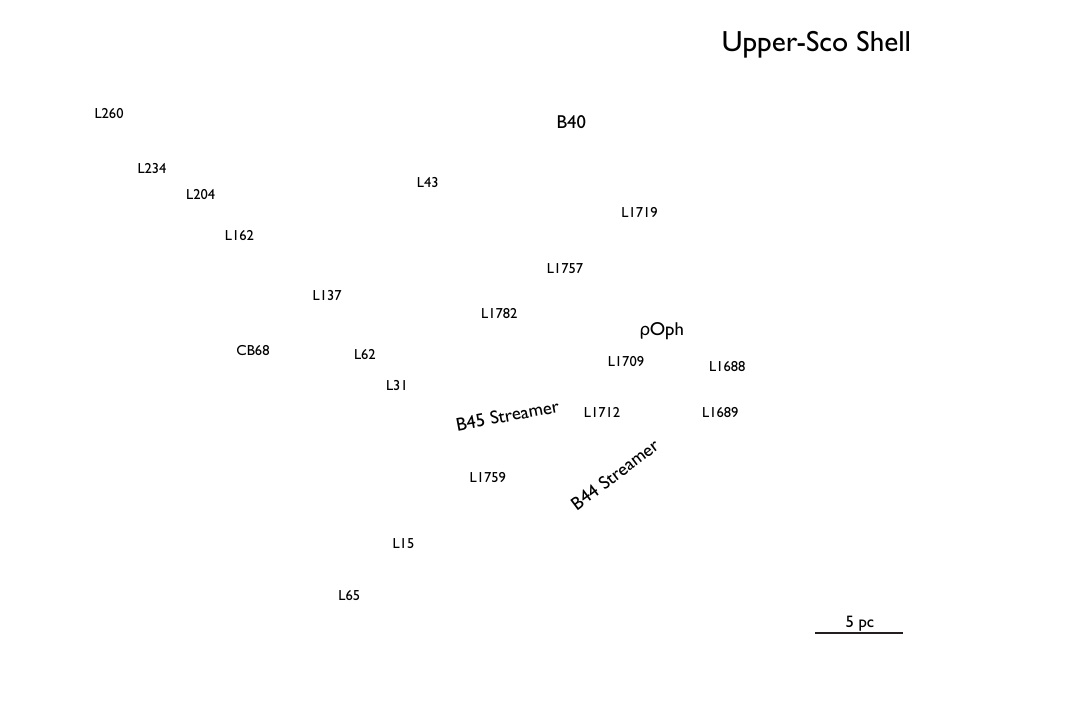}%
  \caption{\textit{Planck-Herschel} optical-depth map for Ophiuchus.  The resolution of the map is 5 arcmin (for the \textit{Planck} data) and 36 arcsec (for the high column density, \textit{Herschel}-covered areas). The corresponding error map is presented in Figure~\ref{fig:PHophmap_Nerr}}
  \label{fig:PHophmap_N}
\end{figure}

\begin{figure}[]
  \centering
  \includegraphics[width=\hsize]{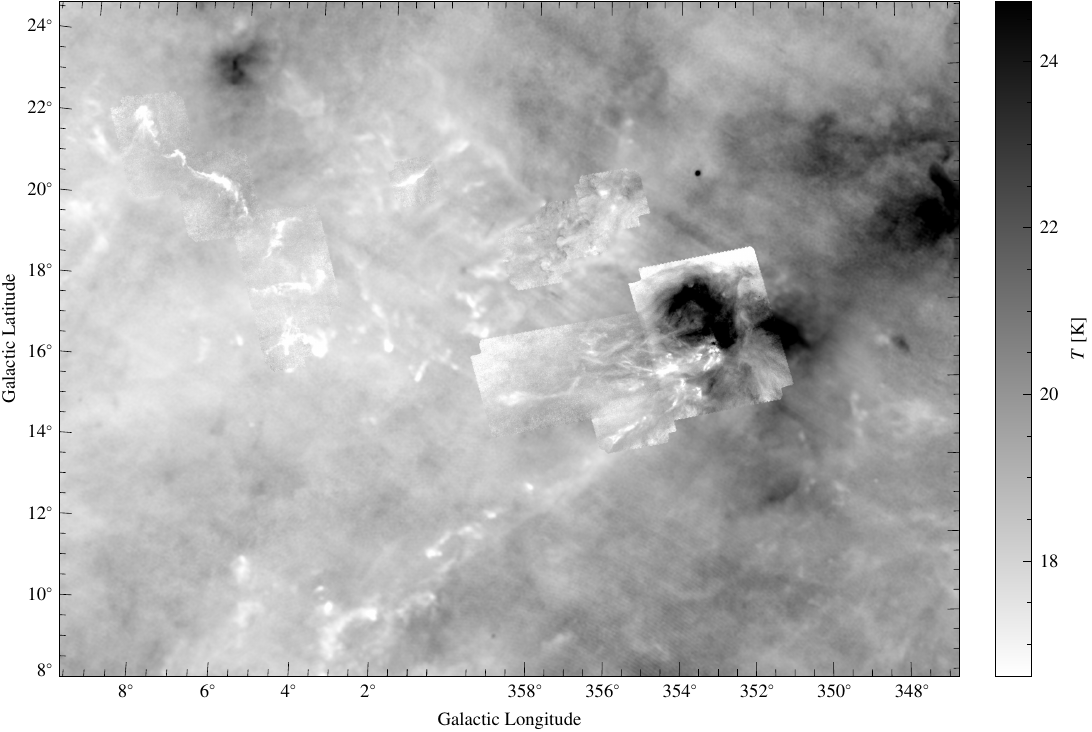}%
  \hspace{-\hsize}%
  \includegraphics[width=\hsize]{Figures/fig03a_labels5.pdf}%
  \caption{Effective dust-temperature map for the Ophiuchus field. The cold structures 
  % (mostly molecular gas, see \cite{Dame2001}) 
  appear white, while the higher temperatures appear as black. There are at least seven hot spots, all associated with massive ionizing stars in the region, implying a distance similar to the cloud material they are warming up. Note, by comparison with Figure~\ref{fig:PHophmap_N}, that most of the hot spots are associated with low column-density regions, the $\rho$ Oph shell being a particularly good example. The corresponding error map is presented in Figure~\ref{fig:PHophmap_Terr}} 
\label{fig:PHophmap_T}
\end{figure}

The interplay between massive ionizing stars and interstellar dust in Figure~\ref{fig:planck3color} is mesmerizing. The color scheme, representing temperatures ranging from about 10 K (in orange) to 30 K (in blue), enables a detailed view of the relationship between radiation from young massive stars and the distribution of warm dust across the star-forming complex. The figure labels the major cloud complexes and notable structures such as the B44 and B45 ``streamers'' and the poorly studied B40 structure \citep{Nozawa1991-id}. The thin solid lines trace stars in the Scorpius constellation, aiding orientation. The Upper-Sco shell \citep{deGeus1992} is indicated by a large dashed circle. A new pillar-like structure near $\pi$ Sco, not detected in the $^{12}$CO survey by \cite{Tachihara2001-wr} but observed in the HI survey by \cite{Kalberla15}, is also highlighted and labeled ``Atomic Pillar''. The existence of a low column-density area, the ``Lupus window,'' near the Lupus 1 cloud \citep{Franco2002-xg}, is confirmed. This figure illustrates the complex interplay between a previous generation of massive stars and star-forming clouds, showing a correlation between heating and gas distribution over tens of parsecs. It highlights the importance of taking into account the wider context of a star-forming complex, rather than concentrating exclusively on particular subregions.

\subsection{Column density and temperature maps}
\label{sec:column-dens-temp}

In Figure~\ref{fig:PHophmap_N} we present the \textit{Planck-Herschel} optical-depth map for the Ophiuchus field constructed using the method described in \cite{Lombardi2014}. The map covers approximately 380 square degrees of the sky ($22.5^{\circ}\times17^{\circ}$) or about $50\times35$ pc$^{2}$ at the distance of the complex. The map has a dynamic range that spans about 3 orders of magnitude, from about $A_{\rm V}\sim0.3$ mag (10$\sigma$) to $A_{\rm V}\sim300$ mag, or about $6\times10^{20}$ to $6\times10^{23}$cm$^{-2}$. This large dynamic range can be reached due to the hybrid approach followed in \cite{Lombardi2014} which combines the higher resolution \textit{Herschel} data with the very low noise \textit{Planck} data. The mean and median column density values on the map are about $A_{\rm V}\sim1.4$ mag. 

In Figure~\ref{fig:PHophmap_T} we present the effective dust temperature map for the combined \textit{Herschel} and \textit{Planck} maps of the Ophiuchus field. The minimum, mean, and maximum temperature on this map are about 11, 19, and 36 K. The temperature maxima in the map coincide with the location of well known ionizing stars, namely the runaway O-star $\zeta$ Oph to the East, $\rho$ Oph (defining an almost circular shell), $\sigma$ Sco, and $\pi$ Sco, which is close to, and likely interacting with, the Atomic pillar structure seen as diffuse blue light in the optical, and as warm dust (about 30 K) in Figure~\ref{fig:planck3color}. 

The combined \textit{Herschel} and \textit{Planck} maps (Figure~\ref{fig:PHophmap_N} and~\ref{fig:PHophmap_T}) provide coverage of the Ophiuchus complex at two different resolutions: 36$^{\prime\prime}$ for the high column density regions covered by \textit{Herschel}, and 5$^\prime$ for the lower density regions covered only by \textit{Planck}. Although the seams between these datasets are generally imperceptible, they become apparent in the L1688 field, which exhibits the highest luminosity. These seams are easily visible on the column density and temperature error maps (Figures~\ref{fig:PHophmap_Nerr} and \ref{fig:PHophmap_Terr}).

In Figure~\ref{fig:Meisner}, we present the \cite{Meisner2014-zn} temperature map of the Ophiuchus-Lupus-Pipe region. \cite{Meisner2014-zn} applied the \cite{Finkbeiner1999-lk} two-component thermal dust emission model to Planck HFI maps for a better fit of the far-infrared dust spectrum and demonstrated that their approach provides more accurate predictions in diffuse sky regions compared to the single-MBB model used by the \cite{Ade2013-no} Planck Collaboration map. The darker grayscales in this figure represent colder temperatures. Compared to the emission map in Figure~\ref{fig:planck3color}, the temperature map provides a better separation between the nearby cold complexes that we want to study and the unrelated background of dust clouds. This is best seen for complexes closer to the Galactic plane, like the Pipe Nebula and some of the Lupus clouds, where the contrast between the nearby cold cloud and the warmer unrelated material along the same line-of-sight is highest. 

\begin{table}[t]
  \caption{Derived masses for cloud structures in Ophiuchus from the HP2 map (Fig.~\ref{fig:PHophmap_N})}             % title of Table
  \label{tab:masses}   
  % is used to refer this table in the text
  \centering                               % used for centering table
\begin{tabular}{lrrrr}
  \hline\hline
    \multicolumn{1}{c}{Name} &
    \multicolumn{1}{c}{Area} &
    \multicolumn{1}{c}{$\langle A_{\rm{K}}\rangle$} &
    \multicolumn{1}{c}{Mass} &
    \multicolumn{1}{c}{Surf. Density} \\
    & pc$^{2}$ & mag & M$_\odot$ & M$_\odot/\rm{pc}^2$\\
  \hline
  Entire HP2 map & 2405.9 & 0.2 & 72660 & 30\\
  $A_{{\rm K}}>0.1$ & 2004.9 & 0.2 & 67048 & 33\\
  $A_{{\rm K}}>0.5$ & 34.9 & 0.9 & 5196 & 149\\
  $A_{{\rm K}}>0.8$ & 10.0 & 1.6 & 2660 & 267\\
  % L1688 & 56.6 & 0.4 & 3940 & 70\\
  L1688 $A_{\rm{K}}>0.5$ & 5.7 & 1.4 & 1367 & 242\\
  L1688 $A_{\rm{K}}>0.8$ & 2.9 & 2.2 & 1081 & 374\\
  % L1689 & 9.5 & 0.6 & 974 & 102\\
  L1689 $A_{\rm{K}}>0.5$ & 4.2 & 1.1 & 780 & 188\\
  L1689 $A_{\rm{K}}>0.8$ & 2.2 & 1.5 & 563 & 261\\
  % B44 & 69.2 & 0.3 & 3742 & 54\\
  B44 $A_{\rm{K}}>0.5$ & 9.9 & 0.9 & 1485 & 150\\
  B44 $A_{\rm{K}}>0.8$ & 6.6 & 1.0 & 1123 & 170\\
  % L1712 & 9.1 & 0.5 & 722 & 79\\
  L1712 $A_{\rm{K}}>0.5$ & 3.3 & 0.8 & 465 & 140\\
  L1712 $A_{\rm{K}}>0.8$ & 1.3 & 1.1 & 250 & 187\\
  % B45 & 20.7 & 0.3 & 1173 & 57\\
  B45 $A_{\rm{K}}>0.5$ & 2.4 & 0.8 & 345 & 141\\
  B45 $A_{\rm{K}}>0.8$ & 0.7 & 1.3 & 147 & 227\\
  % B45 head & 0.5 & 1.5 & 122 & 259\\
  % L1709 & 2.6 & 0.7 & 289 & 111\\
  % L1709 head & 0.3 & 1.4 & 66 & 231\\
  L1709 $A_{\rm{K}}>0.5$ & 1.2 & 1.0 & 206 & 170\\
  L1709 $A_{\rm{K}}>0.8$ & 0.5 & 1.6 & 129 & 279\\
  % L43 & 2.5 & 0.4 & 152 & 60\\
  L43 $A_{\rm{K}}>0.5$ & 0.3 & 1.2 & 52 & 203\\
  L43 $A_{\rm{K}}>0.8$ & 0.2 & 1.5 & 45 & 251\\
  % Oph N & 129.1 & 0.3 & 5705 & 44\\
  Oph N $A_{\rm{K}}>0.5$ & 5.8 & 0.5 & 495 & 85\\
  Oph N $A_{\rm{K}}>0.8$ & 1.7 & 1.3 & 374 & 223\\
  \hline \\
  Oph Arc & 71.9 & 0.2 & 2675 & 37\\
  B40 & 78.0 & 0.3 & 3569 & 46\\
  Atomic Pillar & 3.1 & 0.1 & 55 & 17\\
  \hline\end{tabular}
\end{table}

\subsection{Masses}
\label{sec:masses}

In this section, we estimate the gas masses for the Ophiuchus field (Figure~\ref{fig:PHophmap_N}), and for individual subregions, as a function of a column-density threshold.  To derive masses from the column density map presented in Figure~\ref{fig:PHophmap_N} we integrate the column density over an area of interest in the cloud (as in, e.g., \citealp{Lombardi2014,Zari2015}). In Table~\ref{tab:masses} we present the masses for well-known structures in Ophiuchus, including two distinct column density thresholds ($A_{\rm K} >0.5$ and $A_{\rm K} >0.8$) mag. The former was chosen to separate the main structures as lower thresholds would merge the main structures. The latter threshold is associated with star formation activity \citep{Lada2010} and most of the gas above this threshold is part of only three star-forming gas structures, namely, L1688, B44, and B45. Note that L1689 and L1712, and L1709 are part of the B44 and B45 structures, respectively (see Fig.~\ref{fig:PHophmap_N}). We associate L1719, L1757, and L1782 with B40 to calculate the mass of this structure. We compare our masses estimates to the CO-derived masses in \cite{Loren1989a}, \cite{De_Geus1991-sr}, and the \textit{Herschel}-derived masses in \cite{Howard2021-ur} using a different approach to ours. On average, our masses are about 40\% higher than the CO-derived masses. This suggests that dust emission appears to capture a factor of 1.4 times more mass than CO. Our masses are about 10\% higher than the alternative \textit{Herschel}-derived masses. Given the uncertainties in the definitions of the cloud boundary, in particular for the CO work, the differences might not be substantial. 

We estimate the mass of the entire complex from the Planck map in Figure~\ref{fig:Meisner} to be about $1\times10^5$ M$_\odot$. The total Sco-Cen gas is about $2\times10^5$ M$_\odot$, which means that half of the gas is currently associated with Upper-Sco. To estimate the total mass, we integrated the \cite{Meisner2014-zn} column density map over the area covered by the entire OB association (about $220\times150$ pc$^2$, see Figure 10 of \citealp{Ratzenbock2023-qb}) and subtracted an equal area, equal Galactic longitude, control field taken west of Sco-Cen. We avoided a band of $\pm{3}$ degrees centered on the Galactic plane, where confusion is extreme.

\subsection{Orientation and surface density profile of filaments in complexes surrounding Upper-Sco from Planck data}\label{sec:PlanckEvidence}

\begin{figure*}
  \centering
  \includegraphics[width=\hsize]{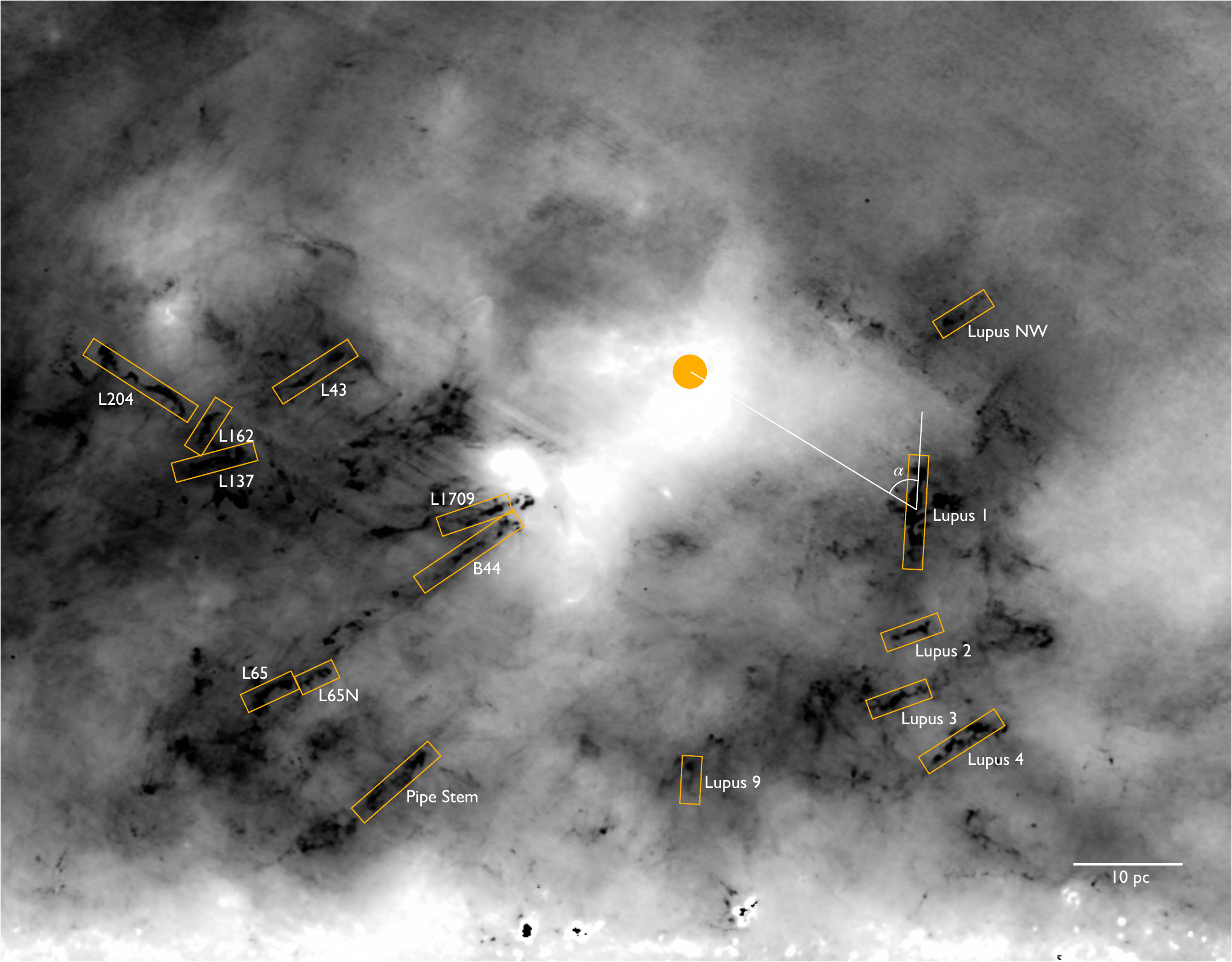}%
  \hspace{-\hsize}%
  \caption{Orientations of the main filaments (enclosed in the orange boxes) in the Ophiuchus-Lupus-Pipe region, shown on the  \cite{Meisner2014-zn} temperature map. The orientation angles were measured between the filament axis and the direction to a central point near the peak of the temperature map, marked with a closed orange circle in the map. This point is consistent with the location of the last supernova in Upper Sco (about 2 Myr ago, \citealt{Neuhauser2019-ck}).}
\label{fig:Meisner}
\end{figure*}

The overall morphology of the region, illustrated in Figure~\ref{fig:Meisner}, including its warmer interior surrounded by colder dense gas arranged in a non-random filament orientations, reveals the influence of feedback from Sco-Cen OB stars. To investigate this potential influence, we analyzed the orientations of the filamentary structures shown in Figure~\ref{fig:Meisner}. We began by estimating the center of Upper-Sco using the distribution of warm dust --- displayed as black (around 13 K) to white regions (around 20 K) in Figure~\ref{fig:Meisner} --- as a tracer of the combined effects of massive stars on the surrounding material. The region of peak temperature aligns with the estimated location of the most recent supernova in Upper-Sco \citep{Neuhauser2019-ck}, and we therefore adopt this supernova's location ($l=-13^\circ$, $b=21^\circ$) as the working center of Upper-Sco.

\begin{figure}
  \centering
  \includegraphics[width=\hsize]{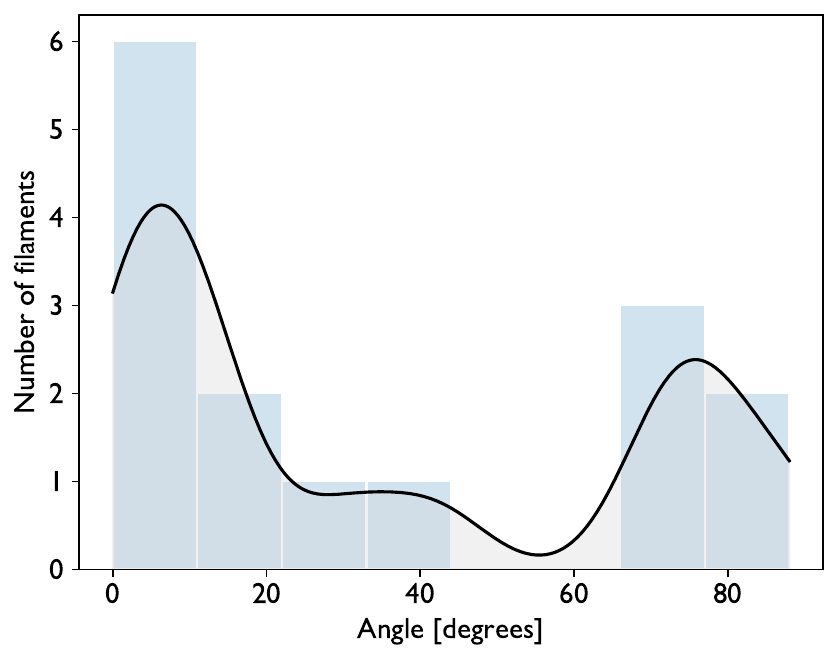}%
  \hspace{\hsize}%
  \caption{Distribution of filament orientations in the Upper-Sco region as shown in Figure~\ref{fig:Meisner}.} 
  % There are two predominant orientations observed: radial and tangential relative to the center of Upper-Sco.}
\label{fig:histogOrientations}
\end{figure}

To determine the orientation of each filament, we measured the angle between two lines: a) the line connecting the adopted center of Upper-Sco to the midpoint of the filament, and (b) the filament's long axis. Figure~\ref{fig:histogOrientations} displays the resulting distribution of filament orientations. Given the improbable assumption that there is a single source of feedback, the distribution exhibits nevertheless two preferred orientations: radial and tangential, with peaks near $0^\circ$ and \textbf{$80^\circ$}, respectively. A Kolmogorov--Smirnov (KS) test confirms that this bimodal distribution is unlikely to occur by chance when compared to a flat distribution ($p=0.03$).

This result is particularly noteworthy because we used a single center for all filaments, which probably does not accurately represent all feedback sources in the region. For example, L204 in Oph North is tangential to, and likely influenced by, the runaway O-star $\zeta$~Oph, located approximately $20^\circ$ away from our adopted center. In this analysis of filament orientation, we focused on the main filamentary structures seen in Figure~\ref{fig:Meisner}, many well-studied molecular clouds. We avoided regions of confusion, without a clear filamentary structure, like B40. Still, many more filamentary structures can be seen through out the map following the general trend of the main filaments. 

\begin{figure*}
  \centering
  \includegraphics[width=\hsize]{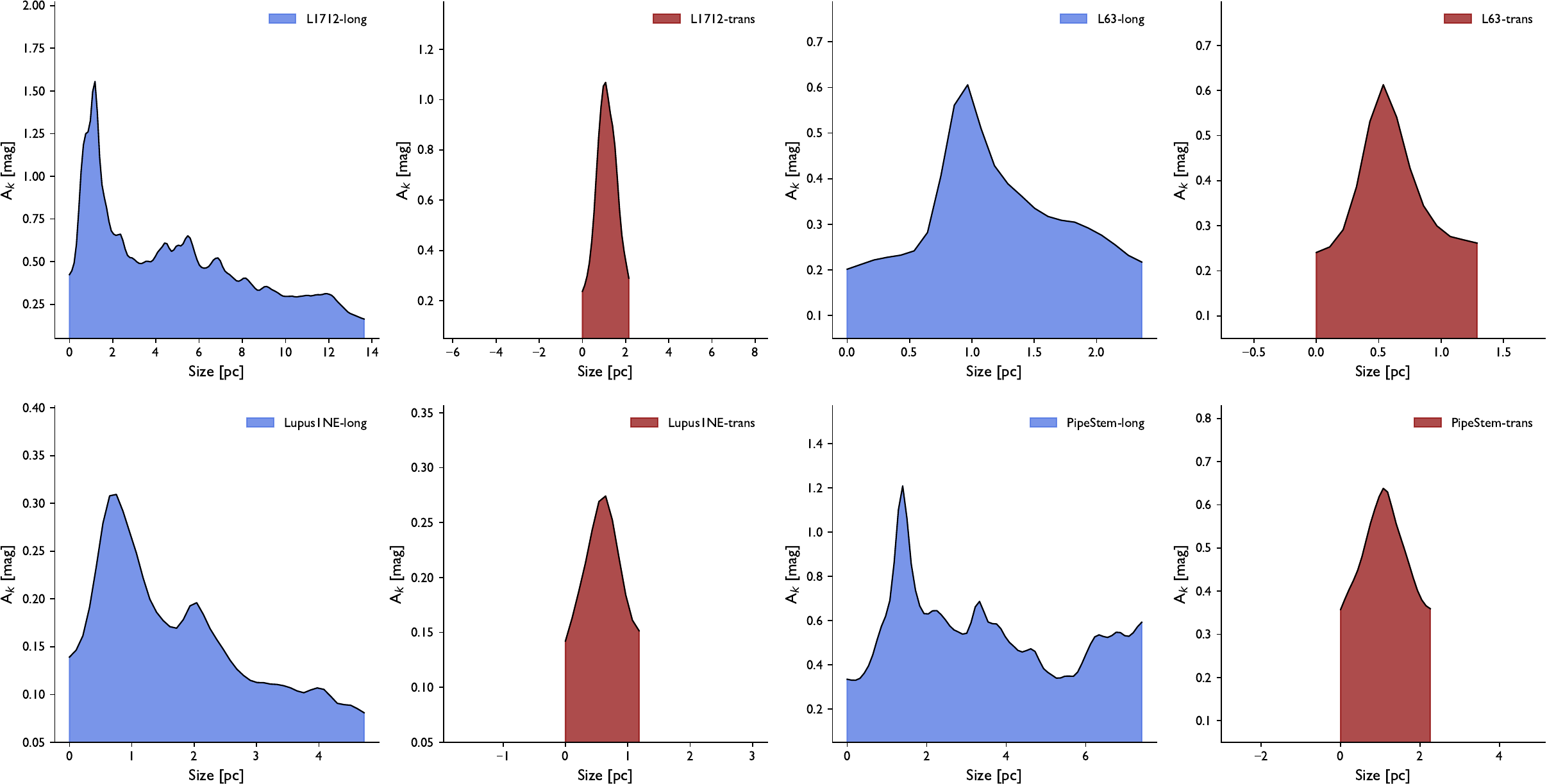}%
  \hspace{-\hsize}%
  \caption{Selected examples of R-type longitudinal (blue) and latitudinal (red) mass profiles. An observable pattern emerges, specifically, R-type filaments predominantly exhibit asymmetric longitudinal distributions (head-tail configuration) and symmetric transverse distributions.}
\label{fig:R-type}
\end{figure*}

We constructed column density profiles for the filaments shown in Figure~\ref{fig:Meisner} using longitudinal (along the filament) and transversal (across the filament) cuts on the Planck column density map derived by \cite{Meisner2014-zn}. Figure~\ref{fig:R-type} illustrates examples of these column density profiles for radial filaments (from hereafter, R-type), shown in blue. 
R-type filaments exhibit asymmetric longitudinal profiles, with their heads oriented towards massive stars. In the larger Upper-Sco region, this pattern holds not only for major Ophiuchus filaments but also for the Pipe Nebula stem and smaller, less-studied filaments. However, the transversal column density profiles of R-type filaments, shown in red, differ significantly, showing approximate symmetry.

\begin{figure*}
  \centering
  \includegraphics[width=\hsize]{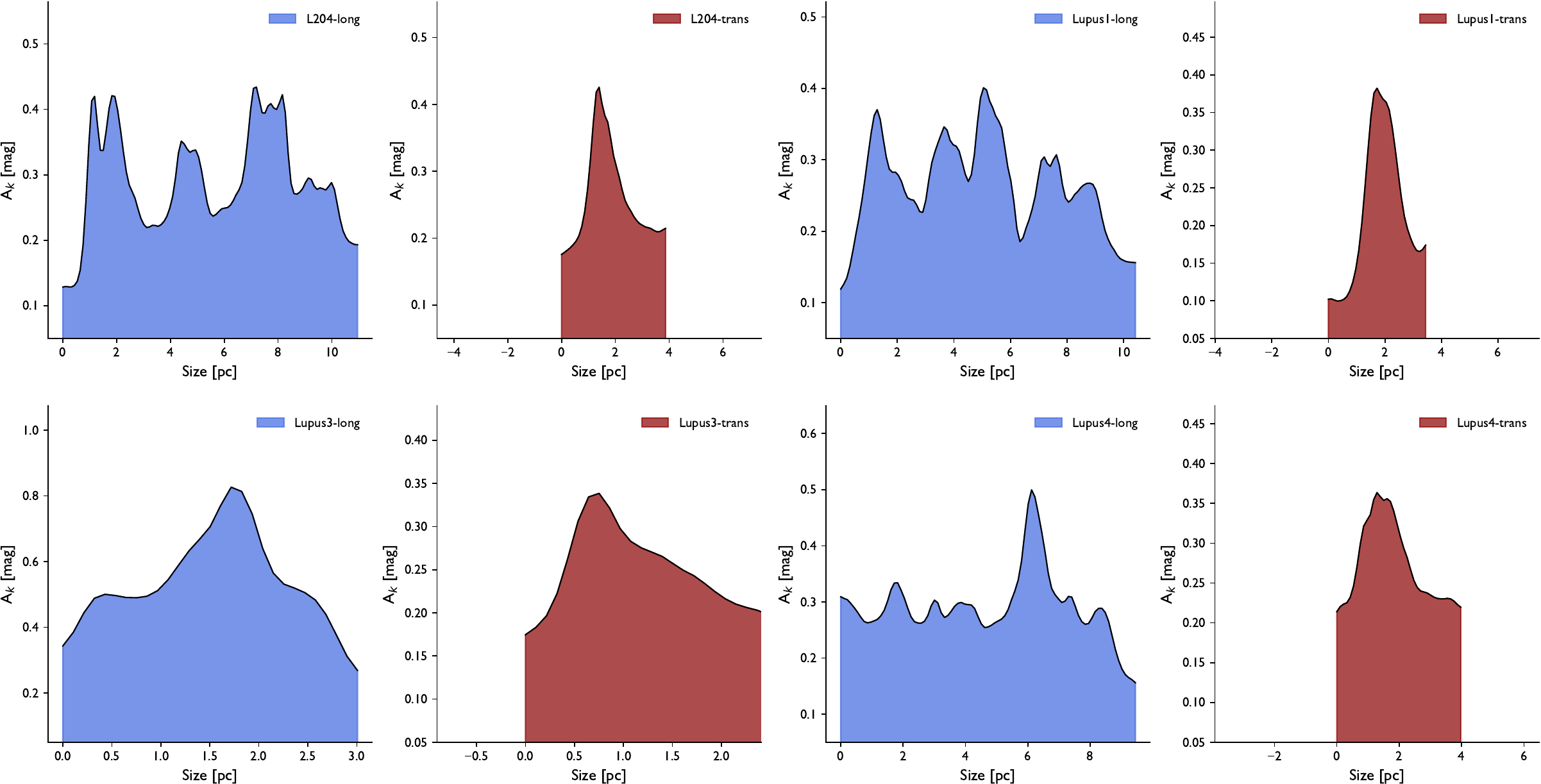}%
  \hspace{-\hsize}%
  \caption{Selected examples of T-type longitudinal (blue) and transversal (red) column density profiles. In contrast to R-type filaments, T-type filaments do not exhibit head-tail morphology. Their longitudinal profiles are generally flat, with dense cores and star formation distributed along their length and no significant mass accumulation toward either end. Interestingly, their transversal profiles often display asymmetry, suggesting mass ``spillover'' on the side facing away from the feedback flow. This asymmetry is another distinguishing feature compared to R-type filaments.
  }
\label{fig:T-type}
\end{figure*}

Figure~\ref{fig:T-type} presents the equivalent analysis for tangential filaments (from hereafter, T-type). T-type longitudinal column density profiles tend to have multiple peaks, contrasting with R-types. Furthermore, T-types lack the distinct longitudinal asymmetry (``head-tail'' morphology) characteristic of R-type filaments.  However, T-type transversal column density profiles are asymmetric, resembling the longitudinal profiles of R-type filaments.


\begin{figure}
  \centering
  \includegraphics[width=\hsize]{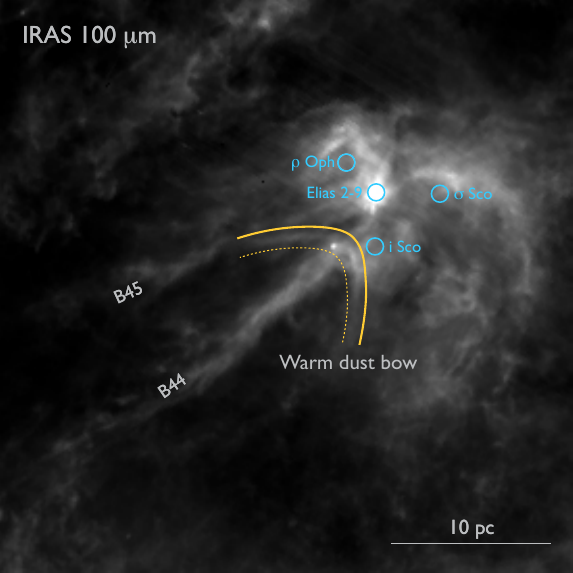}%
  \caption{IRAS 100 $\mu$m image of the Ophiuchus region. The image reveals a bow-like dust feature at the head of B44, suggestive of a bow-shock. If a bow-shock, the Mach number of the flow interacting with B44 is about 2. }
\label{fig:iras}
\end{figure}

\subsection{Further evidence for feedback-driven environment from \textit{Herschel} data}

In this section, we present evidence that feedback from massive stars in Upper-Sco affects the mass distribution of filamentary clouds in Ophiuchus. Although each piece of evidence alone is not definitive, they collectively support this scenario. Several features visible in Figure~\ref{fig:OphClassic} imply a direct influence from massive stars:

\begin{itemize}
    \item  Filamentary structures appear wind-blown, extending away from the massive stars in Upper-Sco (see also Figure~\ref{fig:planck3color})
    \item  The B44 and B45 filaments align approximately radially with the massive stars in Upper-Sco, including $\sigma$ Sco, Elias 2-9 (HD147889), and $\delta$ Sco
    \item Present-day star formation, traced by Class I protostars, occurs within L1688 and at the tips of B44, B45, and L1709 facing the massive stars
    \item  The linear arrangement of Class I protostars in L1688 is orthogonal to the line connecting L1688 and $\sigma$ Sco, suggesting a compressive shock
\end{itemize}

Below, we present complementary evidence for direct interaction between a feedback flow and the gas distribution.

\subsubsection{Bow-shock feature}
Figure~\ref{fig:iras} shows the IRAS 100 $\mu$m image of Ophiuchus, highlighting L1688, B44, B45, and key massive stars. A prominent bow-shaped feature, the ``Warm Dust Bow,'' appears near B44. Faintly detected in \textit{Herschel} data, it likely represents warmer dust. We suggest this is a bow shock formed by interaction between gas at B44’s leading edge and winds from massive stars in Upper-Sco, particularly Elias 2-9 (HD147889).

If this interpretation is correct, the bow shock’s shape provides an estimate of the Mach number $M$. The Mach angle $\theta \approx 35^{\circ}$ implies $M = 1/\sin(\theta) \approx 2$ \citep[e.g.,][]{Landau1987Fluid,shoreAstrophysicalHydrodynamics2007}, indicating a supersonic flow interacting with B44. The shapes of the Oph Arc (near $\sigma$ Sco) and the $\rho$ Oph shell are also consistent with a flow arriving from the northwest, originating near the ionizing stars in Upper-Sco.

\begin{figure}
  \centering
  \includegraphics[width=0.8\hsize]{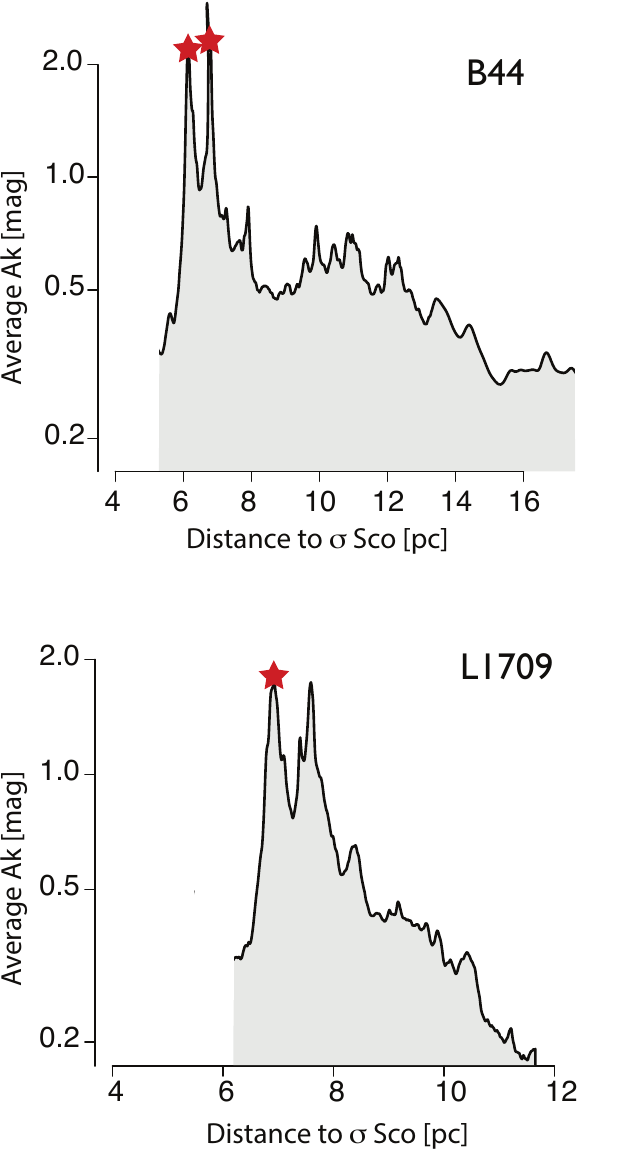}%
  \caption{
  Longitudinal mass profiles of Ophiuchus’s densest R-type filaments (B44 and L1709 at B45’s head) at \textit{Herschel}'s higher resolution. As in the lower resolution Planck-based profile in Figure~\ref{fig:R-type}, the linear mass density decreases from the head, closest to the massive stars, towards the tail. Class I protostars, tracers of ongoing star formation, are shown as red stars and associated with the maxima of the profile. 
  % This gradient supports the idea that feedback from Upper-Sco’s massive stars influences the structure of these filaments.
  } 
\label{fig:profiles}
\end{figure}

\subsubsection{R-type profiles at high resolution} 
Figure~\ref{fig:profiles} shows longitudinal \textit{Herschel} column-density profiles of B44, L1709 (at the head of B45), with Class I protostars marked as red stars. These are higher resolution versions of the Planck-based R-type profiles presented in Figure~\ref{fig:R-type}. The filament's linear mass density decreases from its maximum closer to the massive stars to its minimum downstream from these luminous stars, at the opposite end of the filament. Star formation is occurring only at the maxima of the profile, towards the end of the filament facing the massive stars. The location of ongoing star formation and the filament mass gradient are hard to explain without an interaction between dense gas and a stellar feedback flow. 

\subsubsection{The 3D motion of the Ophiuchus complex}\label{3Dmotion}
Stellar feedback influencing Ophiuchus would imply that it moves away from Upper-Sco’s massive stars under ram pressure. Because most massive stars lie above Ophiuchus, we expect the clouds to move downward if feedback is at play. If Ophiuchus moved upward instead, this would contradict the feedback scenario.

\cite{Grasser2021-tt} analyzed \textit{Gaia} EDR3 data to determine the 3D motion of young stellar objects (YSOs) in Ophiuchus, primarily around L1688. These YSOs, still forming and optically visible, trace the gas motion. They find $UVW = (-5.2 \pm 4.1, -15.1 \pm 1.2, -9.3 \pm 1.6)\,\textrm{km/s}$, differing from Upper-Sco’s $UVW = (-5, -16, -7)\,\textrm{km/s}$ \citep{Luhman2020-ci}. The velocity difference is $\Delta_{UVW} = (0, 0.9, -2)\,\textrm{km/s}$, implying Ophiuchus moves away, mostly downward, from Upper-Sco’s massive stars. This motion supports, but does not prove, the feedback scenario. It also agrees with the ground-based results by \cite{Ducourant2017-qh}, who find $UVW = (-5.9 \pm 0.1, -14.2 \pm 0.3, -8.1 \pm 0.4)\,\textrm{km/s}$.

\cite{Ratzenbock2023-qb,Ratzenbock2023-sw} identified that massive stars such as $\sigma$~Sco and $\delta$~Sco, without \textit{Gaia} parallaxes due to saturation, lie near the centers of newly identified \textit{Gaia} clusters. They likely share similar 3D motions with their clusters. Traceback studies by \cite{Miret-Roig2022-hs} show that the $\rho$ Oph cluster was within \SI{1}{pc} of $\delta$~Sco about \SI{5e6}{yr} ago. This close past interaction suggests that $\delta$~Sco and other massive stars have influenced the Ophiuchus cloud’s morphology and dynamics.

\section{Discussion}
\label{sec:discussion}

\begin{figure}
  \centering
  \includegraphics[width=1\hsize]{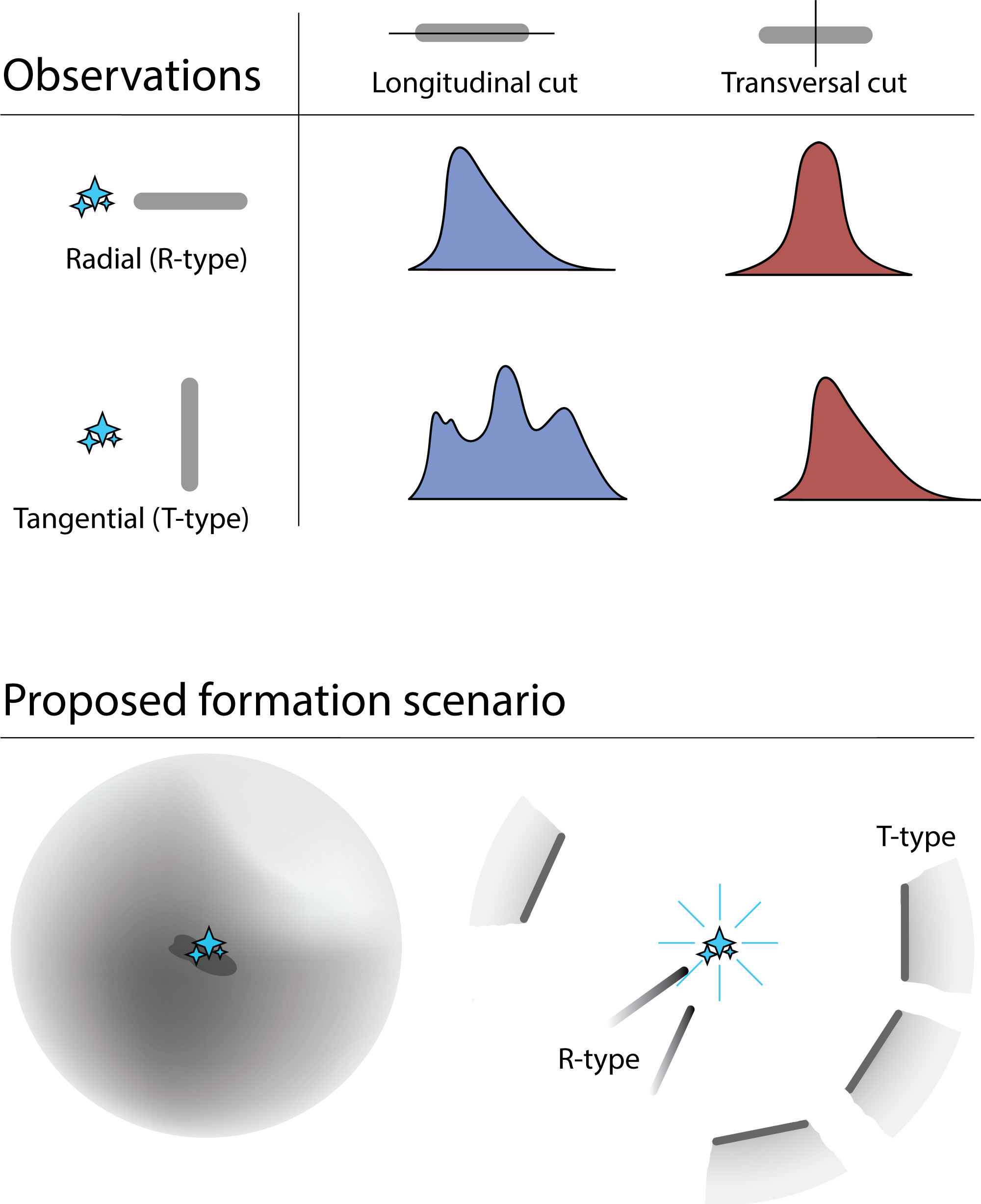}%
  \hspace{-\hsize}%
  \caption{\textbf{Top:} Summary of the longitudinal/transversal profile analysis. R-type profiles exhibit asymmetric longitudinal profiles and symmetric transversal profiles, whereas T-type profiles are characterized by generally flatter longitudinal profiles and asymmetric transversal profiles. These pronounced distinctions between the two filament types imply the existence of two separate mass assembly mechanisms. \textbf{Bottom:} A schematic representation of the proposed formation scenario. Variations in the initial gas distribution following the formation of massive stars may account for the emergence of the two filament types. T-type filaments could arise from a shell fragmentation process, while R-type filaments may result from the stretching and compression of denser gas in proximity to newly formed massive stars. }
\label{fig:CartoonObsModel-type}
\end{figure}

\begin{figure*}
  \centering
  \includegraphics[width=\hsize]{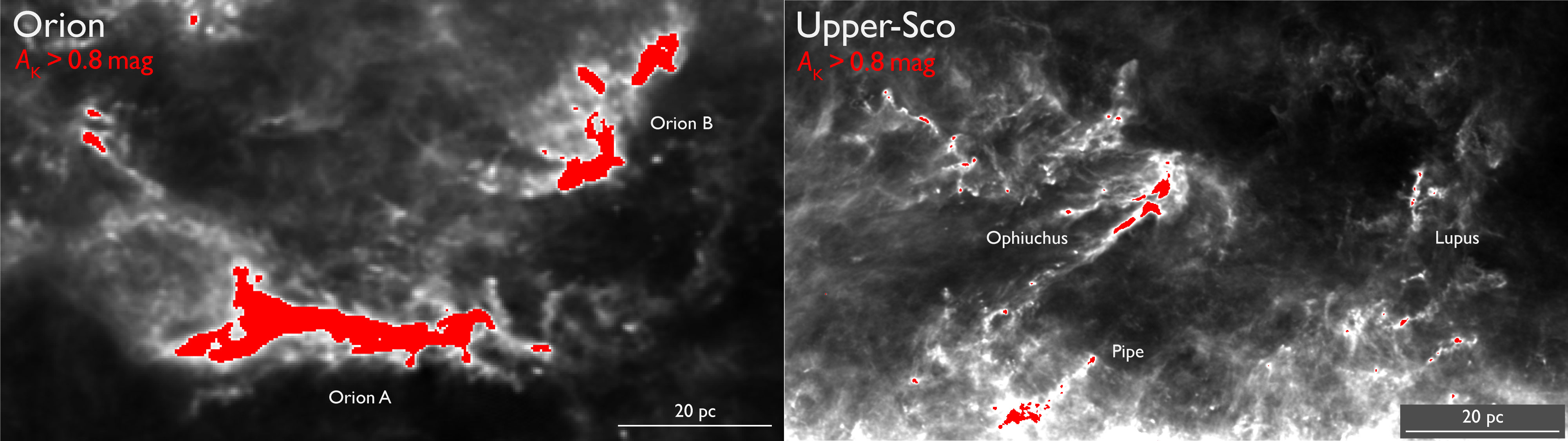}%
  \hspace{-\hsize}%
  \caption{Comparing the Orion and Upper-Sco complexes at the same physical scale reveals striking differences in the amount and distribution of dense gas. The greyscale represents the Planck column-density map \citep{Meisner2014-zn}, and red marks regions with dense gas ($A_K > 0.8$ mag) likely to collapse into new stars. While the gas mass in Orion is about twice that of Upper-Sco, Orion contains ten times more dense gas. Critically, dense gas in Orion is concentrated in two main filamentary clouds, whereas Upper-Sco’s dense gas is more scattered across the region.
  }
\label{fig:Oph-Orion}
\end{figure*}

The arguments presented earlier, and numerous previous studies, support the idea that the feedback from massive stars in Upper-Sco plays a crucial role in shaping the region, aligning with previous studies \citep[e.g.,][]{Vrba1977-me,Loren1986-zj,Loren1989a,deGeus1992,Onishi1999-gd,Tachihara2000-gf,Tachihara2000-iu,Preibisch2008-ln,Peretto2012-jd,Krause2018-vj,Robitaille2018-om,Ladjelate2020-ho}. Furthermore, because of the superior sensitivity and dynamic range of \textit{Planck} and \textit{Herschel}, we found revealing patterns, \emph{birthmarks}, in the alignment of dense gas filaments, their column density profiles, and the locations of star formation activity. All the evidence presented here points to a formation mechanism that organizes gas and star formation in clear patterns over at least 60 pc regions. Following the evidence, we suggest that the feedback flow from massive stars in Upper-Sco is the primary force driving the formation of most filamentary structures in the region. 

The scenario presented here relies on the presence of a significant outflow of diffuse gas from the Sco-Cen association. Until recently, evidence for the existence of such a flow was scarce \citep{Hobbs1969-os,Frisch1986-lv}. This flow was recently reported through ISM absorption lines in this region \citep{Piecka2024-wg}. The outflow has a radial velocity of approximately $-21$ km/s and is generally correlated with the HI emission.  Likely driven by stellar feedback from massive stars in Sco-Cen, the outflow exhibits a complex structure, including a faster component identified through Ca II absorption only. This high-velocity component suggests a dynamic and energetic outflow capable of influencing the surrounding interstellar medium, including the Ophiuchus complex.

\subsection{Creation through destruction: the origin of R-type and T-type filaments}

The simplest model that seems to fit the observations discussed here is one in which a feedback flow from the newly born massive stars towards the center of Upper-Sco rearranges the leftover gas into radial and tangential filaments. An explanation for this bimodal outcome is probably rooted in the initial conditions of the leftover gas. We present two possible formation scenarios below: 

\begin{itemize}
    \item R-type filaments: The flow will quickly disperse the less dense gas leaving behind pockets of compressed denser gas. These will act as a resilient shield against the flow that creates stagnation points, like rocks in a stream \citep[e.g.,][]{Padoan2001,Rogers2013-vx,zamora-avilesStructureExpansionLaw2019}. These dense pockets will cast a radial protective ``shadow'' against the ram pressure of the feedback flow and will naturally develop into head-tail filamentary radial structures \citep[e.g.,][]{Gritschneder2010-ju,Mackey2010-sx}, or R-type filaments with asymmetric longitudinal column density profiles and symmetric transversal profiles (Figure~\ref{fig:R-type}). Unsurprisingly, star formation will mostly take place at these densest regions of the R-type filaments (their heads) and good examples are L1688 ($\rho$-Oph core), L1689, L1709, and B59 in the Pipe Nebula.

An interesting test of this scenario would be to measure the 3D gas flow of any streaming motion along a filament. Although gas kinematic measurements are readily obtained, demonstration of any streaming motion would require knowledge of the filament 3D orientation. Currently, such a measurement is challenging. However, upcoming 3D dust maps of nearby complexes using \textit{Gaia} DR4 data may soon enable measurements of the true space orientation of some filaments. And, in combination with kinematic observations, provide the opportunity to test the feedback-driven formation scenario proposed here.

    \item T-type filaments: The less dense gas, incapable of resisting the ram pressure of the flow, will be pushed outwards and accumulate in a shell at a larger distance from the center. This shell will eventually fragment into filaments that are born tangential to the flow that created them. These tangential filaments, while containing star-forming cores, do not show an obvious gradient in their longitudinal column density profiles (Figure~\ref{fig:T-type}). Interestingly, their asymmetric transversal column density profiles, with excess mass on the side away from the center for most cases, suggests a mass ``spillover'' on the side of the filament opposite the flow \citep{Tachihara2000-iu}, likely due to a process similar to the formation of the head-tail configuration for R-types. The T-type formation mechanism mostly coincides with the classical collect-and-collapse model of \cite{Elmegreen1977} (see also \citealp{Dale2007-da,Zamora-Aviles2019-ry,Whitworth2022-zg}). Good examples of T-type filaments include L204 (tangential to $\zeta$-Oph), Lupus 1, 2, 3, and 4. 
\end{itemize}

\subsection{Timescale considerations}

Can the feedback scenario plausibly explain the formation of a 20 pc R-type filament like B44 within a reasonable timescale? In this scenario, the head of B44 acts as a stagnation point of the flow, resisting it, and suffering the highest flow ram pressure. Flow streamlines would then split and divert around the stagnation point, shearing the denser gas, eventually stretching the original gas cloud into a filament. The feedback scenario then implies that the gas along B44 must have originated, mainly, from the filament head.

A key test for evaluating the feedback scenario in Upper-Sco is the age of the massive stars driving the flow. Based on high-resolution star formation histories derived from \textit{Gaia} data, this age is estimated to be approximately 10 Myr \citep{Ratzenbock2023-sw}. At first approximation, and for R-type filaments, this suggests an average gas streaming velocity of about 2 km/s within the B44 filament, covering the 20 pc distance from head to tail over 10 Myr. This estimate is consistent with the observed radial velocity of the gas across B44 \citep{Loren1989b}, though the true 3D orientation of the filament remains unknown, making the actual streaming velocity uncertain.

Determining the 3D orientation of B44 is crucial for accurately measuring gas motion within the filament, a key aspect of the feedback-driven formation scenario. Confirming the presence of such motion could help identify the dominant feedback processes. For example, \cite{Robitaille2018-om} suggested that both Ophiuchus and Lupus formed as a result of feedback flows triggered by a supernova 1-3 Myr ago. This would imply much higher streaming velocities --- on the order of 7-20 km/s --- which could be possible depending on the actual 3D orientation of B44.

\subsection{A dispersing cloud complex}

With the understanding that Sco-Cen forms a single star formation region containing over $10^5$ M$_\odot$ of gas and approximately 13,000 stars formed in the last 20 Myr, we can place our observations in the broader context of giant molecular cloud (GMC) evolution. This new perspective on the Sco-Cen OB association has been enabled by the exquisite data from ESA's \textit{Gaia} mission \citep{Gaia-Collaboration2023-wg}. \textit{Gaia} revealed a complex tapestry of overlapping young stellar populations and the three-dimensional distribution of gas in the association \citep{Ratzenbock2023-qb,Ratzenbock2023-sw, Edenhofer2023-rp}, providing new insights into the star formation history of Sco-Cen.

The star formation history of Sco-Cen can be summarized as follows: the zenith of star formation activity occurred approximately 15 Myr ago, when the bulk of the stars in the association were formed. Since then, star formation activity has steadily declined to the relatively low levels observed today in Ophiuchus, Lupus, Pipe Nebula, CrA, Chameleon, and L134 \citep{Ratzenbock2023-sw}. These well-studied molecular clouds are the remnants of the primordial Sco-Cen GMC and we are currently witnessing the advanced stages of its dispersal.

To contextualize the Sco-Cen GMC's dispersal, Figure~\ref{fig:Oph-Orion} compares the Orion region (Orion A and B clouds) and the Upper Sco region (Ophiuchus, Lupus, and the Pipe Nebula). Both regions are molecular cloud complexes of comparable mass and size, shown here at the same physical scale for direct comparison. In these maps, grayscale shading represents Planck column density data \citep{Meisner2014-zn}, while regions with $A_K > 0.8$ mag, indicative of dense gas likely to form new stars \citep{Lada2010-tr}, are highlighted in red.
The contrast between the \textit{amount} and \textit{spatial distribution} of dense gas in Orion and Upper-Sco, as illustrated in Figure~\ref{fig:Oph-Orion}, is pronounced. Although Orion contains a total gas mass of $2 \times 10^5$~M$_\odot$ --- about double that of Upper-Sco --- it has ten times more dense gas. Additionally, the dense gas in Orion appears as a cohesive structure concentrated within two major filamentary clouds, whereas in Upper-Sco it is scattered and fragmented across the region, associated with various R- and T-type filaments studied in the previous Section.

Why does Orion contain five times more dense gas per unit mass than Upper-Sco? The distribution of dense gas in Upper-Sco within R- and T-type filaments suggests that feedback from Upper-Sco itself has influenced this arrangement. We propose that the observed differences primarily result from massive star feedback, with Upper-Sco having experienced prolonged exposure to feedback forces relative to Orion. Consider that the main sources of feedback in Upper-Sco are up to 10 Myr and within 10-20 pc of the cloud (see Figure~\ref{fig:planck3color}, while those in a similar proximity to Orion are a few million years old. Figure~\ref{fig:Oph-Orion} thus illustrates the contrasting evolutionary stages of two star-forming GMCs: Orion in an earlier stage of dispersal and Upper-Sco in a later stage.
 
Given the estimated total gas mass in Sco-Cen ($\sim 2\times10^5$ M$_\odot$, see Section~\ref{sec:masses}), the number of Sco-Cen  stars \citep[$\sim13000$,][]{Ratzenbock2023-qb}, and a mean stellar mass of 0.42 Myr \citep{Kroupa2001}, we can estimate the star formation efficiency (SFE) of the large Sco-Cen star forming region to be $\sim2.7\%$ \citep[e.g.,][]{francoMassiveStarformingCapacity1994}. We can also estimate the average Sco-Cen star formation rate (SFR) over the last 20 Myr to be 273 M$_\odot$/Myr, even if Sco-Cen SFR is far from uniform over the last 20 Myr (see Figure 3 in \citealp{Ratzenbock2023-rw}). At the average Sco-Cen SFR, we project that star formation will continue until dense gas consumption in a few Million years, suggesting a Sco-Cen GMC dispersal period of $\leq$ 25 Myr. This dispersal time aligns well with recent estimates of GMC lifetimes, generally in the range of 10-30 Myr \citep{Chevance2022-gx}.

\subsection{The future of the Ophiuchus complex}

We can forecast the future of the Ophiuchus complex, the Upper-Sco gas component containing the largest amount of dense gas in Upper-Sco (see Figure~\ref{fig:Oph-Orion}). The dense gas in Ophiuchus ($A_{\rm K} > 0.8$ mag \citealt{Lada2010-tr}) measured from the \textit{Herschel} column density maps (see Section~\ref{sec:masses}) has the potential to form approximately 800 solar masses of stars. This assumes that most gas at $A_{\rm K} > 0.8$ mag is likely to collapse and form stars at an efficiency of 30\% \citep{Alves2007-he}. This equates to roughly 1900 stars, demonstrating that Ophiuchus remains a fertile region for star formation for the next few Myr, and the most important one in the entire Sco-Cen. The final relative contribution of Ophiuchus to the star formation budget of Sco-Cen is of the order of 10\%. This is a simplified estimate, and it does not account for the potential formation of new dense gas from the compression of diffuse gas.

Rather than forming a single cluster or a random distribution of stars, the future stellar population in Ophiuchus is likely to be structured as the dense gas distribution seen in Figure~\ref{fig:Oph-Orion}. The current configuration of the L1688 core and the main filaments (B44 and B45) resembles the head-tail morphology of the Corona Australis molecular cloud \cite[e.g.,][]{Alves2014-jc}, which is now established to form a chain of young clusters \citep{Ratzenbock2023-sw}. These sequentially aligned clusters exhibit a decreasing age and mass gradient, consistent with a feedback-driven formation scenario \citep{Posch2023-in,Posch2024-mz}. The Ophiuchus complex may give rise to the last chain of clusters in Sco-Cen, and be its star formation closing act.

\section{Conclusions}
\label{sec:conclusions}

We present column density and temperature maps of the entire Ophiuchus cloud complex obtained from \textit{Herschel}, \textit{Planck}, and \textit{2MASS/NICEST} extinction data to establish a comprehensive view of the gas distribution in the complex and look for potential ``birthmarks'' of the filament assembly process. The maps have 36$^{\prime\prime}$ resolution for the regions observed with \textit{Herschel}, and have a dynamic range for the column density covering 0.03 mag to 30 mag of A$_{\rm K}$, or from $6\times10^{20}$ cm$^{-2}$ to $6\times10^{23}$ cm$^{-2}$.

Our primary finding is that stellar feedback from the Upper-Sco region of the Sco-Cen association exerts a dominant influence on the physical structure of gas in the Ophiuchus-Lupus-Pipe region. In this region, we are witnessing close-up of the latter stages of the dispersal of a giant molecular cloud. The feedback, originating from massive stars that began forming approximately 10 Myr ago, shapes the distribution and star formation activity of the leftover gas in the OB association. Specifically, we find that filaments in this region exhibit preferred orientations, tending to align either radially or tangentially with respect to the Upper-Sco massive stars. Furthermore, star formation activity correlates with the spatial distribution of these massive stars. 

The results in this paper illustrate a complex phase in molecular cloud evolution with two simultaneous yet contrasting processes: the formation of filaments and stars via the dispersal of residual gas from a previous massive star formation event. The results also indicate that stellar feedback likely shapes filament formation and evolution in star-forming regions undergoing dispersal by massive stars.

The main results of our analysis can be summarized as follows:

\begin{enumerate}

\item The filaments within the Ophiuchus-Lupus-Pipe region exhibit preferential 
orientations, aligning either radially (R-type) or tangentially (T-type) to massive stars in the region. This suggests that massive star feedback has sculpted the molecular gas, shaping the observed filamentary structures. The age of these filaments is constrained by the age of Upper-Sco, implying they are younger than about 10 Myr. High-resolution \textit{Herschel} data further support this feedback-driven formation scenario.

\item A straightforward scenario, where feedback from the massive stars in Upper-Sco drives a gas flow capable of rearranging and compressing the leftover gas, can, in priciple, explain the observations. Originating from different initial gas distributions, R-type filaments, with star formation at their heads, are likely the remants of denser gas clouds compressed and stretched into filaments, while T-type filaments are the result of a classical collect-and-collapse scenario. These two distinct filament formation mechanisms left observable \emph{``birthmarks''} (orientation, mass distribution, star formation location) on each filament type, providing a testable prediction for future studies of filaments in regions of massive star formation.

\end{enumerate}

From a methodological point of view, our approach highlights the importance of taking into account the wider context of a star-forming complex, rather than concentrating exclusively on particular subregions.

\begin{acknowledgements}

It is a pleasure to acknowledge insightful discussions with Andreas Burkert, Catherine Zucker, Steve Shore, Phil Myers, Alyssa Goodman, Hope Chen, and Juan Soler. Discussions with Julia, Matteo, and Rocco Alves helped in the construction of some of the points made in the paper.  JA acknowledges A. Burkert for the organization of ``Filaments in Molecular Clouds'' in Munich in 2015, and J. Steinacker and A. Bacmann for organizing the \href{https://www2.mpia-hd.mpg.de/homes/stein/EPoS/epos.php}{EPoS} series of meetings where the ideas behind this work were first \href{http://www.mpia.de/homes/stein/EPoS/2016/C/JoaoAlves.php}{presented} and developed. Co-funded by the European Union (ERC, ISM-FLOW, 101055318). Views and opinions expressed are, however, those of the author(s) only and do not necessarily reflect those of the European Union or the European Research Council. Neither the European Union nor the granting authority can be held responsible for them. The results in this paper were based on observations obtained with \textit{Planck} (\url{http://www.esa.int/Planck}), an ESA science mission with instruments and contributions directly funded by ESA Member States, NASA, and Canada. This research has made use of NASA's Astrophysics Data System, the SIMBAD database and the VizieR catalog access tool operated at CDS, Strasbourg, France. This research used Astropy, a community-developed core Python package for Astronomy \citep{2013A&A...558A..33A} and TOPCAT, an interactive graphical viewer and editor for tabular data \citep{2005ASPC..347...29T}.
\end{acknowledgements}

\bibliographystyle{aa} 

\bibliography{zmy}%,zmy}

\begin{thebibliography}{132}
\expandafter\ifx\csname natexlab\endcsname\relax\def\natexlab#1{#1}\fi

\bibitem[{{Abergel} {et~al.}(1994){Abergel}, {Boulanger}, {Mizuno}, \& {Fukui}}]{Abergel1994}
{Abergel}, A., {Boulanger}, F., {Mizuno}, A., \& {Fukui}, Y. 1994, \apjl, 423, L59

\bibitem[{Ade {et~al.}(2014)Ade, Aghanim, Alves, {Armitage-Caplan}, Arnaud, Ashdown, {Atrio-Barandela}, Aumont, Aussel, Baccigalupi, Banday, Barreiro, Barrena, Bartelmann, Bartlett, Bartolo, Basak, Battaner, Battye, Benabed, Beno{\^i}t, {Benoit-L{\'e}vy}, Bernard, Bersanelli, Bertincourt, Bethermin, Bielewicz, Bikmaev, Blanchard, Bobin, Bock, B{\"o}hringer, Bonaldi, Bonavera, Bond, Borrill, Bouchet, Boulanger, Bourdin, Bowyer, Bridges, Brown, Bucher, Burenin, Burigana, Butler, Calabrese, Cappellini, Cardoso, Carr, Carvalho, Casale, Castex, Catalano, Challinor, Chamballu, Chary, Chen, Chiang, Chiang, Chon, Christensen, Churazov, Church, Clemens, Clements, Colombi, Colombo, Combet, Comis, Couchot, Coulais, Crill, Cruz, Curto, Cuttaia, Silva, Dahle, Danese, Davies, Davis, de~Bernardis, de~Rosa, de~Zotti, D{\'e}chelette, Delabrouille, Delouis, D{\'e}mocl{\`e}s, D{\'e}sert, Dick, Dickinson, Diego, Dolag, Dole, Donzelli, Dor{\'e}, Douspis, Ducout, Dunkley, Dupac, Efstathiou, Elsner, En{\ss}lin, Eriksen, Fabre, Falgarone, Falvella, Fantaye, Fergusson, Filliard, Finelli, {Flores-Cacho}, Foley, Forni, Fosalba, Frailis, Fraisse, Franceschi, Freschi, Fromenteau, Frommert, Gaier, Galeotta, Gallegos, Galli, Gandolfo, Ganga, Gauthier, {G{\'e}nova-Santos}, Ghosh, Giard, Giardino, Gilfanov, Girard, {Giraud-H{\'e}raud}, Gjerl{\o}w, {Gonz{\'a}lez-Nuevo}, G{\'o}rski, Gratton, Gregorio, Gruppuso, Gudmundsson, Haissinski, Hamann, Hansen, Hansen, Hanson, Harrison, Heavens, Helou, Hempel, {Henrot-Versill{\'e}}, {Hern{\'a}ndez-Monteagudo}, Herranz, Hildebrandt, Hivon, Ho, Hobson, Holmes, Hornstrup, Hou, Hovest, Huey, Huffenberger, Hurier, Ili{\'c}, Jaffe, Jaffe, Jasche, Jewell, Jones, Juvela, Kalberla, Kangaslahti, Keih{\"a}nen, Kerp, Keskitalo, Khamitov, Kiiveri, Kim, Kisner, Kneissl, Knoche, Knox, Kunz, {Kurki-Suonio}, Lacasa, Lagache, L{\"a}hteenm{\"a}ki, Lamarre, Langer, Lasenby, Lattanzi, Laureijs, Lavabre, Lawrence, Jeune, Leach, Leahy, Leonardi, {Le{\'o}n-Tavares}, Leroy, Lesgourgues, Lewis, Li, Liddle, Liguori, Lilje, {Linden-V{\o}rnle}, Lindholm, {L{\'o}pez-Caniego}, Lowe, Lubin, {Mac{\'i}as-P{\'e}rez}, MacTavish, Maffei, Maggio, Maino, Mandolesi, Mangilli, {Marcos-Caballero}, Marinucci, Maris, Marleau, Marshall, Martin, {Mart{\'i}nez-Gonz{\'a}lez}, Masi, Massardi, Matarrese, Matsumura, Matthai, Maurin, Mazzotta, McDonald, McEwen, McGehee, Mei, Meinhold, Melchiorri, Melin, Mendes, Menegoni, Mennella, Migliaccio, Mikkelsen, Millea, Miniscalco, Mitra, {Miville-Desch{\^e}nes}, Molinari, Moneti, Montier, Morgante, Morisset, Mortlock, Moss, Munshi, Murphy, Naselsky, Nati, Natoli, Negrello, Nesvadba, Netterfield, {N{\o}rgaard-Nielsen}, North, Noviello, Novikov, Novikov, O'Dwyer, Orieux, Osborne, O'Sullivan, Oxborrow, Paci, Pagano, Pajot, Paladini, Pandolfi, Paoletti, Partridge, Pasian, Patanchon, Paykari, Pearson, Pearson, Peel, Peiris, Perdereau, Perotto, Perrotta, Pettorino, Piacentini, Piat, Pierpaoli, Pietrobon, Plaszczynski, Platania, Pogosyan, Pointecouteau, Polenta, Ponthieu, Popa, Poutanen, Pratt, Pr{\'e}zeau, Prunet, Puget, Pullen, Rachen, Racine, Rahlin, R{\"a}th, Reach, Rebolo, Reinecke, Remazeilles, Renault, Renzi, Riazuelo, Ricciardi, Riller, Ringeval, Ristorcelli, Robbers, Rocha, Roman, Rosset, Rossetti, Roudier, {Rowan-Robinson}, {Rubi{\~n}o-Mart{\'i}n}, {Ruiz-Granados}, Rusholme, Salerno, Sandri, Sanselme, Santos, Savelainen, Savini, Schaefer, Schiavon, Scott, Seiffert, Serra, Shellard, Smith, Smoot, Souradeep, Spencer, Starck, Stolyarov, Stompor, Sudiwala, Sunyaev, Sureau, Sutter, Sutton, {Suur-Uski}, Sygnet, Tauber, Tavagnacco, Taylor, Terenzi, Texier, Toffolatti, Tomasi, Torre, Tristram, Tucci, Tuovinen, T{\"u}rler, Tuttlebee, Umana, Valenziano, Valiviita, Tent, Varis, Vibert, Viel, Vielva, Villa, Vittorio, Wade, Wandelt, Watson, Watson, Wehus, Welikala, Weller, White, White, Wilkinson, Winkel, Xia, Yvon, Zacchei, Zibin, \& Zonca}]{2013arXiv1303.5062P}
Ade, P. a.~R., Aghanim, N., Alves, M. I.~R., {et~al.} 2014, Astronomy \& Astrophysics, 571, A1

\bibitem[{Ade {et~al.}(2013)Ade, Aghanim, Alves, Arnaud, Ashdown, Atrio-Barandela, Aumont, Baccigalupi, Balbi, Banday, Barreiro, Bartlett, Battaner, Bedini, Benabed, Benoît, Bernard, Bersanelli, Bonaldi, Bond, Borrill, Bouchet, Boulanger, Burigana, Butler, Cabella, Cardoso, Chen, Chiang, Christensen, Clements, Colombi, Colombo, Coulais, Cuttaia, Davies, Davis, de~Bernardis, de~Gasperis, de~Zotti, Delabrouille, Dickinson, Diego, Dobler, Dole, Donzelli, Doré, Douspis, Dupac, Enßlin, Finelli, Forni, Frailis, Franceschi, Galeotta, Ganga, Génova-Santos, Ghosh, Giard, Giardino, Giraud-Héraud, González-Nuevo, Górski, Gregorio, Gruppuso, Hansen, Harrison, Hernández-Monteagudo, Hildebrandt, Hivon, Hobson, Holmes, Hornstrup, Hovest, Huffenberger, Jaffe, Jaffe, Juvela, Keihänen, Keskitalo, Kisner, Knoche, Kunz, Kurki-Suonio, Lagache, Lähteenmäki, Lamarre, Lasenby, Lawrence, Leach, Leonardi, Lilje, Linden-Vørnle, Lubin, Macías-Pérez, Maffei, Maino, Mandolesi, Maris, Marshall, Martin, Martínez-González, Masi, Massardi, Matarrese, Mazzotta, Melchiorri, Mennella, Mitra, Miville-Deschênes, Moneti, Montier, Morgante, Mortlock, Munshi, Murphy, Naselsky, Nati, Natoli, Nørgaard-Nielsen, Noviello, Novikov, Novikov, Osborne, Oxborrow, Pajot, Paladini, Paoletti, Peel, Perotto, Perrotta, Piacentini, Piat, Pierpaoli, Pietrobon, Plaszczynski, Pointecouteau, Polenta, Popa, Poutanen, Pratt, Prunet, Puget, Rachen, Reach, Rebolo, Reinecke, Renault, Ricciardi, Ristorcelli, Rocha, Rosset, Rubiño-Martín, Rusholme, Salerno, Sandri, Savini, Scott, Spencer, Stolyarov, Sudiwala, Suur-Uski, Sygnet, Tauber, Terenzi, Tibbs, Toffolatti, Tomasi, Tristram, Valenziano, Van~Tent, Varis, Vielva, Villa, Vittorio, Wade, Wandelt, Ysard, Yvon, Zacchei, \& Zonca}]{Ade2013-no}
Ade, P. A.~R., Aghanim, N., Alves, M. I.~R., {et~al.} 2013, A\&A, 557, A53

\bibitem[{{Alves} {et~al.}(1998){Alves}, {Lada}, {Lada}, {Kenyon}, \& {Phelps}}]{Alves1998}
{Alves}, J., {Lada}, C.~J., {Lada}, E.~A., {Kenyon}, S.~J., \& {Phelps}, R. 1998, \apj, 506, 292

\bibitem[{Alves {et~al.}(2007)Alves, Lombardi, \& Lada}]{Alves2007-he}
Alves, J., Lombardi, M., \& Lada, C. 2007, in ISLAND UNIVERSES, Astrophysics and Space Science Proceedings (Springer, Dordrecht), 417--422

\bibitem[{Alves {et~al.}(2014)Alves, Lombardi, \& Lada}]{Alves2014-jc}
Alves, J., Lombardi, M., \& Lada, C.~J. 2014, A\&A, 565, A18

\bibitem[{{Andr{\'e}} {et~al.}(2010){Andr{\'e}}, {Men'shchikov}, {Bontemps}, {K{\"o}nyves}, {Motte}, {Schneider}, {Didelon}, {Minier}, {Saraceno}, {Ward-Thompson}, {di Francesco}, {White}, {Molinari}, {Testi}, {Abergel}, {Griffin}, {Henning}, {Royer}, {Mer{\'{\i}}n}, {Vavrek}, {Attard}, {Arzoumanian}, {Wilson}, {Ade}, {Aussel}, {Baluteau}, {Benedettini}, {Bernard}, {Blommaert}, {Cambr{\'e}sy}, {Cox}, {di Giorgio}, {Hargrave}, {Hennemann}, {Huang}, {Kirk}, {Krause}, {Launhardt}, {Leeks}, {Le Pennec}, {Li}, {Martin}, {Maury}, {Olofsson}, {Omont}, {Peretto}, {Pezzuto}, {Prusti}, {Roussel}, {Russeil}, {Sauvage}, {Sibthorpe}, {Sicilia-Aguilar}, {Spinoglio}, {Waelkens}, {Woodcraft}, \& {Zavagno}}]{Andre2010}
{Andr{\'e}}, P., {Men'shchikov}, A., {Bontemps}, S., {et~al.} 2010, \aap, 518, L102

\bibitem[{{Andre} \& {Montmerle}(1994)}]{Andre1994}
{Andre}, P. \& {Montmerle}, T. 1994, \apj, 420, 837

\bibitem[{Arzoumanian {et~al.}(2018)Arzoumanian, Shimajiri, Inutsuka, Inoue, \& Tachihara}]{Arzoumanian2018-fj}
Arzoumanian, D., Shimajiri, Y., Inutsuka, S.-I., Inoue, T., \& Tachihara, K. 2018, PASJ, 70

\bibitem[{{Bailer-Jones}(2015)}]{Bailer-Jones2015}
{Bailer-Jones}, C.~A.~L. 2015, \pasp, 127, 994

\bibitem[{{Bally} {et~al.}(1987){Bally}, {Langer}, {Stark}, \& {Wilson}}]{Bally1987}
{Bally}, J., {Langer}, W.~D., {Stark}, A.~A., \& {Wilson}, R.~W. 1987, \apjl, 312, L45

\bibitem[{{Barnard}(1907)}]{Barnard1907}
{Barnard}, E.~E. 1907, \apj, 25

\bibitem[{{Barnard} {et~al.}(1927){Barnard}, {Frost}, \& {Calvert}}]{barnard1927z}
{Barnard}, E.~E., {Frost}, E.~B., \& {Calvert}, M.~R. 1927, {A Photographic Atlas of Selected Regions of the Milky Way}

\bibitem[{{Blaauw}(1964)}]{Blaauw1964}
{Blaauw}, A. 1964, \araa, 2, 213

\bibitem[{Bonne {et~al.}(2020)Bonne, Bontemps, Schneider, Clarke, Arzoumanian, Fukui, Tachihara, Csengeri, Guesten, Ohama, Okamoto, Simon, Yahia, \& Yamamoto}]{Bonne2020-we}
Bonne, L., Bontemps, S., Schneider, N., {et~al.} 2020, A\&A, 644, A27

\bibitem[{{Boulanger} {et~al.}(1985){Boulanger}, {Baud}, \& {van Albada}}]{Boulanger1985}
{Boulanger}, F., {Baud}, B., \& {van Albada}, G.~D. 1985, \aap, 144, L9

\bibitem[{Bouy \& Alves(2015)}]{Bouy2015-ce}
Bouy, H. \& Alves, J. 2015, A\&A, 584, A26

\bibitem[{Burkert \& Hartmann(2004)}]{Burkert2004-np}
Burkert, A. \& Hartmann, L. 2004, ApJ, 616, 288

\bibitem[{{Cernicharo} {et~al.}(1985){Cernicharo}, {Bachiller}, \& {Duvert}}]{Cernicharo1985}
{Cernicharo}, J., {Bachiller}, R., \& {Duvert}, G. 1985, \aap, 149, 273

\bibitem[{Chevance {et~al.}(2022)Chevance, Krumholz, McLeod, Ostriker, Rosolowsky, \& Sternberg}]{Chevance2022-gx}
Chevance, M., Krumholz, M.~R., McLeod, A.~F., {et~al.} 2022

\bibitem[{Dale {et~al.}(2007)Dale, Bonnell, \& Whitworth}]{Dale2007-da}
Dale, J.~E., Bonnell, I.~A., \& Whitworth, A.~P. 2007, MNRAS, 375, 1291–1298

\bibitem[{{de Geus}(1992)}]{deGeus1992}
{de Geus}, E.~J. 1992, \aap, 262, 258

\bibitem[{de~Geus \& Burton(1991)}]{De_Geus1991-sr}
de~Geus, E.~J. \& Burton, W.~B. 1991, A\&A, 246, 559

\bibitem[{Ducourant {et~al.}(2017)Ducourant, Teixeira, Krone-Martins, Bontemps, Despois, Galli, Bouy, Le~Campion, Rapaport, \& Cuillandre}]{Ducourant2017-qh}
Ducourant, C., Teixeira, R., Krone-Martins, A., {et~al.} 2017, A\&A, 597, A90

\bibitem[{Edenhofer {et~al.}(2024)Edenhofer, Zucker, Frank, Saydjari, Speagle, Finkbeiner, \& En{\ss}lin}]{Edenhofer2023-rp}
Edenhofer, G., Zucker, C., Frank, P., {et~al.} 2024, A\&A, 685, A82

\bibitem[{{Elmegreen} \& {Lada}(1977)}]{Elmegreen1977}
{Elmegreen}, B.~G. \& {Lada}, C.~J. 1977, \apj, 214, 725

\bibitem[{Finkbeiner {et~al.}(1999)Finkbeiner, Davis, \& Schlegel}]{Finkbeiner1999-lk}
Finkbeiner, D.~P., Davis, M., \& Schlegel, D.~J. 1999, ApJ, 524, 867

\bibitem[{Franco(2002)}]{Franco2002-xg}
Franco, G. A.~P. 2002, MNRAS, 331, 474

\bibitem[{Franco {et~al.}(1994)Franco, Shore, \& {Tenorio-Tagle}}]{francoMassiveStarformingCapacity1994}
Franco, J., Shore, S.~N., \& {Tenorio-Tagle}, G. 1994, ApJ, 436, 795

\bibitem[{Frisch \& York(1986)}]{Frisch1986-lv}
Frisch, P. \& York, D. 1986, in The galaxy and the solar system (A87-34101 14-90). Tucson, AZ, University of Arizona Press, 1986, p. 83-100

\bibitem[{{Gaia Collaboration} {et~al.}(2023){Gaia Collaboration}, Vallenari, Brown, Prusti, de~Bruijne, Arenou, Babusiaux, Biermann, Creevey, Ducourant, Evans, Eyer, Guerra, Hutton, Jordi, Klioner, Lammers, Lindegren, Luri, Mignard, Panem, Pourbaix, Randich, Sartoretti, Soubiran, Tanga, Walton, Bailer-Jones, Bastian, Drimmel, Jansen, Katz, Lattanzi, van Leeuwen, Bakker, Cacciari, Castañeda, De~Angeli, Fabricius, Fouesneau, Frémat, Galluccio, Guerrier, Heiter, Masana, Messineo, Mowlavi, Nicolas, Nienartowicz, Pailler, Panuzzo, Riclet, Roux, Seabroke, Sordo, Thévenin, Gracia-Abril, Portell, Teyssier, Altmann, Andrae, Audard, Bellas-Velidis, Benson, Berthier, Blomme, Burgess, Busonero, Busso, Cánovas, Carry, Cellino, Cheek, Clementini, Damerdji, Davidson, de~Teodoro, Nuñez~Campos, Delchambre, Dell'Oro, Esquej, Fernández-Hernández, Fraile, Garabato, García-Lario, Gosset, Haigron, Halbwachs, Hambly, Harrison, Hernández, Hestroffer, Hodgkin, Holl, Janßen, Jevardat~de Fombelle, Jordan, Krone-Martins, Lanzafame, Löffler, Marchal, Marrese, Moitinho, Muinonen, Osborne, Pancino, Pauwels, Recio-Blanco, Reylé, Riello, Rimoldini, Roegiers, Rybizki, Sarro, Siopis, Smith, Sozzetti, Utrilla, van Leeuwen, Abbas, Ábrahám, Abreu~Aramburu, Aerts, Aguado, Ajaj, Aldea-Montero, Altavilla, Álvarez, Alves, Anders, Anderson, Anglada~Varela, Antoja, Baines, Baker, Balaguer-Núñez, Balbinot, Balog, Barache, Barbato, Barros, Barstow, Bartolomé, Bassilana, Bauchet, Becciani, Bellazzini, Berihuete, Bernet, Bertone, Bianchi, Binnenfeld, Blanco-Cuaresma, Blazere, Boch, Bombrun, Bossini, Bouquillon, Bragaglia, Bramante, Breedt, Bressan, Brouillet, Brugaletta, Bucciarelli, Burlacu, Butkevich, Buzzi, Caffau, Cancelliere, Cantat-Gaudin, Carballo, Carlucci, Carnerero, Carrasco, Casamiquela, Castellani, Castro-Ginard, Chaoul, Charlot, Chemin, Chiaramida, Chiavassa, Chornay, Comoretto, Contursi, Cooper, Cornez, Cowell, Crifo, Cropper, Crosta, Crowley, Dafonte, Dapergolas, David, David, de~Laverny, De~Luise, De~March, De~Ridder, de~Souza, de~Torres, del Peloso, del Pozo, Delbo, Delgado, Delisle, Demouchy, Dharmawardena, Di~Matteo, Diakite, Diener, Distefano, Dolding, Edvardsson, Enke, Fabre, Fabrizio, Faigler, Fedorets, Fernique, Fienga, Figueras, Fournier, Fouron, Fragkoudi, Gai, Garcia-Gutierrez, Garcia-Reinaldos, García-Torres, Garofalo, Gavel, Gavras, Gerlach, Geyer, Giacobbe, Gilmore, Girona, Giuffrida, Gomel, Gomez, González-Núñez, González-Santamaría, González-Vidal, Granvik, Guillout, Guiraud, Gutiérrez-Sánchez, Guy, Hatzidimitriou, Hauser, Haywood, Helmer, Helmi, Sarmiento, Hidalgo, Hilger, Hładczuk, Hobbs, Holland, Huckle, Jardine, Jasniewicz, Jean-Antoine~Piccolo, Jiménez-Arranz, Jorissen, Juaristi~Campillo, Julbe, Karbevska, Kervella, Khanna, Kontizas, Kordopatis, Korn, Kóspál, Kostrzewa-Rutkowska, Kruszyńska, Kun, Laizeau, Lambert, Lanza, Lasne, Le~Campion, Lebreton, Lebzelter, Leccia, Leclerc, Lecoeur-Taibi, Liao, Licata, Lindstrøm, Lister, Livanou, Lobel, Lorca, Loup, Madrero~Pardo, Magdaleno~Romeo, Managau, Mann, Manteiga, Marchant, Marconi, Marcos, Marcos~Santos, Marín~Pina, Marinoni, Marocco, Marshall, Martin~Polo, Martín-Fleitas, Marton, Mary, Masip, Massari, Mastrobuono-Battisti, Mazeh, McMillan, Messina, Michalik, Millar, Mints, Molina, Molinaro, Molnár, Monari, Monguió, Montegriffo, Montero, Mor, Mora, Morbidelli, Morel, Morris, Muraveva, Murphy, Musella, Nagy, Noval, Ocaña, Ogden, Ordenovic, Osinde, Pagani, Pagano, Palaversa, Palicio, Pallas-Quintela, Panahi, Payne-Wardenaar, Peñalosa~Esteller, Penttilä, Pichon, Piersimoni, Pineau, Plachy, Plum, Poggio, Prša, Pulone, Racero, Ragaini, Rainer, Raiteri, Rambaux, Ramos, Ramos-Lerate, Re~Fiorentin, Regibo, Richards, Rios~Diaz, Ripepi, Riva, Rix, Rixon, Robichon, Robin, Robin, Roelens, Rogues, Rohrbasser, Romero-Gómez, Rowell, Royer, Ruz~Mieres, Rybicki, Sadowski, Sáez~Núñez, Sagristà~Sellés, Sahlmann, Salguero, Samaras, Sanchez~Gimenez, Sanna, Santoveña, Sarasso, Schultheis, Sciacca, Segol, Segovia, Ségransan, Semeux, Shahaf, Siddiqui, Siebert, Siltala, Silvelo, Slezak, Slezak, Smart, Snaith, Solano, Solitro, Souami, Souchay, Spagna, Spina, Spoto, Steele, Steidelmüller, Stephenson, Süveges, Surdej, Szabados, Szegedi-Elek, Taris, Taylor, Teixeira, Tolomei, Tonello, Torra, Torra, Torralba~Elipe, Trabucchi, Tsounis, Turon, Ulla, Unger, Vaillant, van Dillen, van Reeven, Vanel, Vecchiato, Viala, Vicente, Voutsinas, Weiler, Wevers, Wyrzykowski, Yoldas, Yvard, Zhao, Zorec, Zucker, \& Zwitter}]{Gaia-Collaboration2023-wg}
{Gaia Collaboration}, Vallenari, A., Brown, A. G.~A., {et~al.} 2023, A\&A, 674, A1

\bibitem[{{Goldsmith} {et~al.}(2008){Goldsmith}, {Heyer}, {Narayanan}, {Snell}, {Li}, \& {Brunt}}]{Goldsmith2008}
{Goldsmith}, P.~F., {Heyer}, M., {Narayanan}, G., {et~al.} 2008, \apj, 680, 428

\bibitem[{Grasser {et~al.}(2021)Grasser, Ratzenböck, Alves, Großschedl, Meingast, Zucker, Hacar, Lada, Goodman, Lombardi, Forbes, Bomze, \& Möller}]{Grasser2021-tt}
Grasser, N., Ratzenböck, S., Alves, J., {et~al.} 2021, A\&A, 652, A2

\bibitem[{{Griffin} {et~al.}(2010){Griffin}, {Abergel}, {Abreu}, {Ade}, {Andr{\'e}}, {Augueres}, {Babbedge}, {Bae}, {Baillie}, {Baluteau}, {Barlow}, {Bendo}, {Benielli}, {Bock}, {Bonhomme}, {Brisbin}, {Brockley-Blatt}, {Caldwell}, {Cara}, {Castro-Rodriguez}, {Cerulli}, {Chanial}, {Chen}, {Clark}, {Clements}, {Clerc}, {Coker}, {Communal}, {Conversi}, {Cox}, {Crumb}, {Cunningham}, {Daly}, {Davis}, {de Antoni}, {Delderfield}, {Devin}, {di Giorgio}, {Didschuns}, {Dohlen}, {Donati}, {Dowell}, {Dowell}, {Duband}, {Dumaye}, {Emery}, {Ferlet}, {Ferrand}, {Fontignie}, {Fox}, {Franceschini}, {Frerking}, {Fulton}, {Garcia}, {Gastaud}, {Gear}, {Glenn}, {Goizel}, {Griffin}, {Grundy}, {Guest}, {Guillemet}, {Hargrave}, {Harwit}, {Hastings}, {Hatziminaoglou}, {Herman}, {Hinde}, {Hristov}, {Huang}, {Imhof}, {Isaak}, {Israelsson}, {Ivison}, {Jennings}, {Kiernan}, {King}, {Lange}, {Latter}, {Laurent}, {Laurent}, {Leeks}, {Lellouch}, {Levenson}, {Li}, {Li}, {Lilienthal}, {Lim}, {Liu}, {Lu}, {Madden}, {Mainetti}, {Marliani}, {McKay}, {Mercier}, {Molinari}, {Morris}, {Moseley}, {Mulder}, {Mur}, {Naylor}, {Nguyen}, {O'Halloran}, {Oliver}, {Olofsson}, {Olofsson}, {Orfei}, {Page}, {Pain}, {Panuzzo}, {Papageorgiou}, {Parks}, {Parr-Burman}, {Pearce}, {Pearson}, {P{\'e}rez-Fournon}, {Pinsard}, {Pisano}, {Podosek}, {Pohlen}, {Polehampton}, {Pouliquen}, {Rigopoulou}, {Rizzo}, {Roseboom}, {Roussel}, {Rowan-Robinson}, {Rownd}, {Saraceno}, {Sauvage}, {Savage}, {Savini}, {Sawyer}, {Scharmberg}, {Schmitt}, {Schneider}, {Schulz}, {Schwartz}, {Shafer}, {Shupe}, {Sibthorpe}, {Sidher}, {Smith}, {Smith}, {Smith}, {Spencer}, {Stobie}, {Sudiwala}, {Sukhatme}, {Surace}, {Stevens}, {Swinyard}, {Trichas}, {Tourette}, {Triou}, {Tseng}, {Tucker}, {Turner}, {Vaccari}, {Valtchanov}, {Vigroux}, {Virique}, {Voellmer}, {Walker}, {Ward}, {Waskett}, {Weilert}, {Wesson}, {White}, {Whitehouse}, {Wilson}, {Winter}, {Woodcraft}, {Wright}, {Xu}, {Zavagno}, {Zemcov}, {Zhang}, \& {Zonca}}]{Griffin2010}
{Griffin}, M.~J., {Abergel}, A., {Abreu}, A., {et~al.} 2010, \aap, 518, L3

\bibitem[{Gritschneder {et~al.}(2010)Gritschneder, Burkert, Naab, \& Walch}]{Gritschneder2010-ju}
Gritschneder, M., Burkert, A., Naab, T., \& Walch, S. 2010, ApJ, 723, 971

\bibitem[{Hacar {et~al.}(2023)Hacar, Clark, Heitsch, Kainulainen, Panopoulou, Seifried, \& Smith}]{Hacar2022-gd}
Hacar, A., Clark, S.~E., Heitsch, F., {et~al.} 2023, in PPVII

\bibitem[{Hacar {et~al.}(2018)Hacar, Tafalla, Forbrich, Alves, Meingast, Grossschedl, \& Teixeira}]{Hacar2018-bt}
Hacar, A., Tafalla, M., Forbrich, J., {et~al.} 2018, A\&A, 610, A77

\bibitem[{{Haffner} {et~al.}(2003){Haffner}, {Reynolds}, {Tufte}, {Madsen}, {Jaehnig}, \& {Percival}}]{Haffner2003}
{Haffner}, L.~M., {Reynolds}, R.~J., {Tufte}, S.~L., {et~al.} 2003, \apjs, 149, 405

\bibitem[{{Hatchell} {et~al.}(2012){Hatchell}, {Terebey}, {Huard}, {Mamajek}, {Allen}, {Bourke}, {Dunham}, {Gutermuth}, {Harvey}, {J{\o}rgensen}, {Mer{\'{\i}}n}, {Noriega-Crespo}, \& {Peterson}}]{Hatchell12}
{Hatchell}, J., {Terebey}, S., {Huard}, T., {et~al.} 2012, \apj, 754, 104

\bibitem[{{Heiles} \& {Jenkins}(1976)}]{Heiles1976}
{Heiles}, C. \& {Jenkins}, E.~B. 1976, \aap, 46, 333

\bibitem[{Heitsch(2013)}]{Heitsch2013-we}
Heitsch, F. 2013, 769, 115

\bibitem[{{Hennebelle}(2013)}]{Hennebelle2013}
{Hennebelle}, P. 2013, \aap, 556, A153

\bibitem[{Hobbs(1969)}]{Hobbs1969-os}
Hobbs, L.~M. 1969, ApJ, 158, 461

\bibitem[{Howard {et~al.}(2021)Howard, Whitworth, Griffin, Marsh, \& Smith}]{Howard2021-ur}
Howard, A. D.~P., Whitworth, A.~P., Griffin, M.~J., Marsh, K.~A., \& Smith, M. W.~L. 2021, MNRAS, 504, 6157

\bibitem[{Inutsuka {et~al.}(2015)Inutsuka, Inoue, Iwasaki, \& Hosokawa}]{Inutsuka2015-io}
Inutsuka, S.-I., Inoue, T., Iwasaki, K., \& Hosokawa, T. 2015, A\&A, 580, A49

\bibitem[{Jeffreson {et~al.}(2020)Jeffreson, Kruijssen, Keller, Chevance, \& Glover}]{Jeffreson2020-om}
Jeffreson, S. M.~R., Kruijssen, J. M.~D., Keller, B.~W., Chevance, M., \& Glover, S. C.~O. 2020, MNRAS, 498, 385

\bibitem[{{Johnstone} \& {Bally}(1999)}]{Johnstone1999}
{Johnstone}, D. \& {Bally}, J. 1999, \apjl, 510, L49

\bibitem[{{Kalberla} \& {Haud}(2015)}]{Kalberla15}
{Kalberla}, P.~M.~W. \& {Haud}, U. 2015, \aap, 578, A78

\bibitem[{Krause {et~al.}(2018)Krause, Burkert, Diehl, Fierlinger, Gaczkowski, Kroell, Ngoumou, Roccatagliata, Siegert, \& Preibisch}]{Krause2018-vj}
Krause, M. G.~H., Burkert, A., Diehl, R., {et~al.} 2018, A\&A, 619, A120

\bibitem[{{Kroupa}(2001)}]{Kroupa2001}
{Kroupa}, P. 2001, \mnras, 322, 231

\bibitem[{{Kryukova} {et~al.}(2012){Kryukova}, {Megeath}, {Gutermuth}, {Pipher}, {Allen}, {Allen}, {Myers}, \& {Muzerolle}}]{Kryukova2012}
{Kryukova}, E., {Megeath}, S.~T., {Gutermuth}, R.~A., {et~al.} 2012, \aj, 144, 31

\bibitem[{{Lada} {et~al.}(1994){Lada}, {Lada}, {Clemens}, \& {Bally}}]{Lada1994}
{Lada}, C.~J., {Lada}, E.~A., {Clemens}, D.~P., \& {Bally}, J. 1994, \apj, 429, 694

\bibitem[{{Lada} {et~al.}(2010){Lada}, {Lombardi}, \& {Alves}}]{Lada2010}
{Lada}, C.~J., {Lombardi}, M., \& {Alves}, J.~F. 2010, \apj, 724, 687

\bibitem[{Lada {et~al.}(2010)Lada, Lombardi, \& Alves}]{Lada2010-tr}
Lada, C.~J., Lombardi, M., \& Alves, J.~F. 2010, ApJ

\bibitem[{{Lada} \& {Wilking}(1984)}]{Lada1984}
{Lada}, C.~J. \& {Wilking}, B.~A. 1984, \apj, 287, 610

\bibitem[{Ladjelate {et~al.}(2020)Ladjelate, André, Könyves, Ward-Thompson, Men'shchikov, Bracco, Palmeirim, Roy, Shimajiri, Kirk, Arzoumanian, Benedettini, Di~Francesco, Fiorellino, Schneider, Pezzuto, Motte, \& {the Herschel Gould Belt Survey Team}}]{Ladjelate2020-ho}
Ladjelate, B., André, P., Könyves, V., {et~al.} 2020, A\&A, 638, A74

\bibitem[{Landau \& Lifshitz(1987)}]{Landau1987Fluid}
Landau, L.~D. \& Lifshitz, E.~M. 1987, Fluid Mechanics, Second Edition: Volume 6 (Course of Theoretical Physics), Course of theoretical physics / by L. D. Landau and E. M. Lifshitz, Vol. 6 (Butterworth-Heinemann)

\bibitem[{Liseau {et~al.}(2015)Liseau, Larsson, Lunttila, Olberg, Rydbeck, Bergman, Justtanont, Olofsson, \& de~Vries}]{Liseau2015-zp}
Liseau, R., Larsson, B., Lunttila, T., {et~al.} 2015, Gas and dust in the star-forming regionrho Oph A

\bibitem[{{Loinard} {et~al.}(2008){Loinard}, {Torres}, {Mioduszewski}, \& {Rodr{\'{\i}}guez}}]{Loinard2008}
{Loinard}, L., {Torres}, R.~M., {Mioduszewski}, A.~J., \& {Rodr{\'{\i}}guez}, L.~F. 2008, \apjl, 675, L29

\bibitem[{{Lombardi} {et~al.}(2006){Lombardi}, {Alves}, \& {Lada}}]{Lombardi2006}
{Lombardi}, M., {Alves}, J., \& {Lada}, C.~J. 2006, \aap, 454, 781

\bibitem[{Lombardi {et~al.}(2011)Lombardi, Alves, \& Lada}]{Lombardi2011-ae}
Lombardi, M., Alves, J., \& Lada, C.~J. 2011, A\&A, 535, A16

\bibitem[{{Lombardi} {et~al.}(2014){Lombardi}, {Bouy}, {Alves}, \& {Lada}}]{Lombardi2014}
{Lombardi}, M., {Bouy}, H., {Alves}, J., \& {Lada}, C.~J. 2014, \aap, 566, A45

\bibitem[{Lombardi {et~al.}(2008)Lombardi, Lada, \& Alves}]{Lombardi2008-sa}
Lombardi, M., Lada, C.~J., \& Alves, J. 2008, A\&A, 489, 143

\bibitem[{{Lombardi} {et~al.}(2008){Lombardi}, {Lada}, \& {Alves}}]{Lombardi2008}
{Lombardi}, M., {Lada}, C.~J., \& {Alves}, J. 2008, \aap, 489, 143

\bibitem[{Lombardi {et~al.}(2008)Lombardi, Lada, \& Alves}]{Lombardi2008-qi}
Lombardi, M., Lada, C.~J., \& Alves, J. 2008, A\&A, 480, 785

\bibitem[{{Lombardi} {et~al.}(2010){Lombardi}, {Lada}, \& {Alves}}]{Lombardi2010}
{Lombardi}, M., {Lada}, C.~J., \& {Alves}, J. 2010, \aap, 512, 67

\bibitem[{{Loren}(1989{\natexlab{a}})}]{Loren1989a}
{Loren}, R.~B. 1989{\natexlab{a}}, \apj, 338, 902

\bibitem[{{Loren}(1989{\natexlab{b}})}]{Loren1989b}
{Loren}, R.~B. 1989{\natexlab{b}}, \apj, 338, 925

\bibitem[{Loren \& Wootten(1986)}]{Loren1986-zj}
Loren, R.~B. \& Wootten, A. 1986, ApJ, 306, 142

\bibitem[{Luhman \& Esplin(2020)}]{Luhman2020-ci}
Luhman, K.~L. \& Esplin, T.~L. 2020, AJ, 160

\bibitem[{{Lynds}(1962)}]{Lynds1962}
{Lynds}, B.~T. 1962, \apjs, 7, 1

\bibitem[{Mackey {et~al.}(2015)Mackey, Gvaramadze, Mohamed, \& Langer}]{Mackey2015}
Mackey, J., Gvaramadze, V.~V., Mohamed, S., \& Langer, N. 2015, A\&A, 573, A10

\bibitem[{Mackey \& Lim(2010)}]{Mackey2010-sx}
Mackey, J. \& Lim, A.~J. 2010, MNRAS, 403, 714–730

\bibitem[{{McClure-Griffiths} {et~al.}(2006){McClure-Griffiths}, {Dickey}, {Gaensler}, {Green}, \& {Haverkorn}}]{McClure-Griffiths2006}
{McClure-Griffiths}, N.~M., {Dickey}, J.~M., {Gaensler}, B.~M., {Green}, A.~J., \& {Haverkorn}, M. 2006, \apj, 652, 1339

\bibitem[{Meisner \& Finkbeiner(2014)}]{Meisner2014-zn}
Meisner, A.~M. \& Finkbeiner, D.~P. 2014, ApJ, 798, 88

\bibitem[{Miret-Roig {et~al.}(2022)Miret-Roig, Galli, Olivares, Bouy, Alves, \& Barrado}]{Miret-Roig2022-hs}
Miret-Roig, N., Galli, P. A.~B., Olivares, J., {et~al.} 2022, A\&A, 667, A163

\bibitem[{{Miville-Desch{\^e}nes} {et~al.}(2010){Miville-Desch{\^e}nes}, {Martin}, {Abergel}, {Bernard}, {Boulanger}, {Lagache}, {Anderson}, {Andr{\'e}}, {Arab}, {Baluteau}, {Blagrave}, {Bontemps}, {Cohen}, {Compiegne}, {Cox}, {Dartois}, {Davis}, {Emery}, {Fulton}, {Gry}, {Habart}, {Huang}, {Joblin}, {Jones}, {Kirk}, {Lim}, {Madden}, {Makiwa}, {Menshchikov}, {Molinari}, {Moseley}, {Motte}, {Naylor}, {Okumura}, {Pinheiro Gon{\c c}alves}, {Polehampton}, {Rod{\'o}n}, {Russeil}, {Saraceno}, {Schneider}, {Sidher}, {Spencer}, {Swinyard}, {Ward-Thompson}, {White}, \& {Zavagno}}]{Miville-Desch2010}
{Miville-Desch{\^e}nes}, M.-A., {Martin}, P.~G., {Abergel}, A., {et~al.} 2010, \aap, 518, L104

\bibitem[{{Molinari} {et~al.}(2010){Molinari}, {Swinyard}, {Bally}, {Barlow}, {Bernard}, {Martin}, {Moore}, {Noriega-Crespo}, {Plume}, {Testi}, {Zavagno}, {Abergel}, {Ali}, {Andr{\'e}}, {Baluteau}, {Benedettini}, {Bern{\'e}}, {Billot}, {Blommaert}, {Bontemps}, {Boulanger}, {Brand}, {Brunt}, {Burton}, {Campeggio}, {Carey}, {Caselli}, {Cesaroni}, {Cernicharo}, {Chakrabarti}, {Chrysostomou}, {Codella}, {Cohen}, {Compiegne}, {Davis}, {de Bernardis}, {de Gasperis}, {Di Francesco}, {di Giorgio}, {Elia}, {Faustini}, {Fischera}, {Fukui}, {Fuller}, {Ganga}, {Garcia-Lario}, {Giard}, {Giardino}, {Glenn}, {Goldsmith}, {Griffin}, {Hoare}, {Huang}, {Jiang}, {Joblin}, {Joncas}, {Juvela}, {Kirk}, {Lagache}, {Li}, {Lim}, {Lord}, {Lucas}, {Maiolo}, {Marengo}, {Marshall}, {Masi}, {Massi}, {Matsuura}, {Meny}, {Minier}, {Miville-Desch{\^e}nes}, {Montier}, {Motte}, {M{\"u}ller}, {Natoli}, {Neves}, {Olmi}, {Paladini}, {Paradis}, {Pestalozzi}, {Pezzuto}, {Piacentini}, {Pomar{\`e}s}, {Popescu}, {Reach}, {Richer}, {Ristorcelli}, {Roy}, {Royer}, {Russeil}, {Saraceno}, {Sauvage}, {Schilke}, {Schneider-Bontemps}, {Schuller}, {Schultz}, {Shepherd}, {Sibthorpe}, {Smith}, {Smith}, {Spinoglio}, {Stamatellos}, {Strafella}, {Stringfellow}, {Sturm}, {Taylor}, {Thompson}, {Tuffs}, {Umana}, {Valenziano}, {Vavrek}, {Viti}, {Waelkens}, {Ward-Thompson}, {White}, {Wyrowski}, {Yorke}, \& {Zhang}}]{Molinari2010}
{Molinari}, S., {Swinyard}, B., {Bally}, J., {et~al.} 2010, \pasp, 122, 314

\bibitem[{{Montmerle} {et~al.}(1983){Montmerle}, {Koch-Miramond}, {Falgarone}, \& {Grindlay}}]{Montmerle1983}
{Montmerle}, T., {Koch-Miramond}, L., {Falgarone}, E., \& {Grindlay}, J.~E. 1983, \apj, 269, 182

\bibitem[{{Myers}(2009)}]{Myers2009}
{Myers}, P.~C. 2009, \apj, 700, 1609

\bibitem[{{Nagai} {et~al.}(1998){Nagai}, {Inutsuka}, \& {Miyama}}]{Nagai1998}
{Nagai}, T., {Inutsuka}, S.-i., \& {Miyama}, S.~M. 1998, \apj, 506, 306

\bibitem[{Neuhäuser {et~al.}(2019)Neuhäuser, Gießler, \& Hambaryan}]{Neuhauser2019-ck}
Neuhäuser, R., Gießler, F., \& Hambaryan, V.~V. 2019, MNRAS

\bibitem[{{North} {et~al.}(2007){North}, {Davis}, {Tuthill}, {Tango}, \& {Robertson}}]{North2007}
{North}, J.~R., {Davis}, J., {Tuthill}, P.~G., {Tango}, W.~J., \& {Robertson}, J.~G. 2007, \mnras, 380, 1276

\bibitem[{North {et~al.}(2007)North, Davis, Tuthill, Tango, \& Robertson}]{North2007-tt}
North, J.~R., Davis, J., Tuthill, P.~G., Tango, W.~J., \& Robertson, J.~G. 2007, MNRAS, 380, 1276

\bibitem[{Nozawa {et~al.}(1991)Nozawa, Mizuno, Teshima, Ogawa, \& Fukui}]{Nozawa1991-id}
Nozawa, S., Mizuno, A., Teshima, Y., Ogawa, H., \& Fukui, Y. 1991, ApJS, 77, 647

\bibitem[{Nutter {et~al.}(2006)Nutter, Ward-Thompson, \& André}]{Nutter2006-he}
Nutter, D., Ward-Thompson, D., \& André, P. 2006, MNRAS, 368, 1833

\bibitem[{{Ochsendorf} {et~al.}(2014){Ochsendorf}, {Cox}, {Krijt}, {Salgado}, {Bern{\'e}}, {Bernard}, {Kaper}, \& {Tielens}}]{Ochsendorf2014}
{Ochsendorf}, B.~B., {Cox}, N.~L.~J., {Krijt}, S., {et~al.} 2014, \aap, 563, A65

\bibitem[{Onishi {et~al.}(1999)Onishi, Kawamura, Abe, Yamaguchi, Saito, Moriguchi, Mizuno, Ogawa, \& Fukui}]{Onishi1999-gd}
Onishi, T., Kawamura, A., Abe, R., {et~al.} 1999, PASJ, 51, 871

\bibitem[{Ortiz-León {et~al.}(2017)Ortiz-León, Loinard, Kounkel, Dzib, Mioduszewski, Rodríguez, Torres, González-Lópezlira, Pech, Rivera, Hartmann, Boden, Evans, Briceño, Tobin, Galli, \& Gudehus}]{Ortiz-Leon2017-em}
Ortiz-León, G.~N., Loinard, L., Kounkel, M.~A., {et~al.} 2017, ApJ, 834, 141

\bibitem[{{Ott}(2010)}]{2010ASPC..434..139O}
{Ott}, S. 2010, in ASPC, Vol. 434, Astronomical Data Analysis Software and Systems XIX, ed. Y.~{Mizumoto}, K.-I. {Morita}, \& M.~{Ohishi}, 139

\bibitem[{{Padoan} {et~al.}(2001){Padoan}, {Juvela}, {Goodman}, \& {Nordlund}}]{Padoan2001}
{Padoan}, P., {Juvela}, M., {Goodman}, A.~A., \& {Nordlund}, {\AA}. 2001, \apj, 553, 227

\bibitem[{Peretto {et~al.}(2012)Peretto, André, Könyves, Schneider, Arzoumanian, Palmeirim, Didelon, Attard, Bernard, Di~Francesco, Elia, Hennemann, Hill, Kirk, Men'shchikov, Motte, Nguyen~Luong, Roussel, Sousbie, Testi, Ward-Thompson, White, \& Zavagno}]{Peretto2012-jd}
Peretto, N., André, P., Könyves, V., {et~al.} 2012, A\&A, 541, A63

\bibitem[{Piecka {et~al.}(2024)Piecka, Hutschenreuter, \& Alves}]{Piecka2024-wg}
Piecka, M., Hutschenreuter, S., \& Alves, J. 2024, A\&A, 689, A84

\bibitem[{Pillsworth \& Pudritz(2024)}]{Pillsworth2024-sd}
Pillsworth, R. \& Pudritz, R.~E. 2024, Mon. Not. R. Astron. Soc., 528, 209

\bibitem[{Pineda {et~al.}(2023)Pineda, Arzoumanian, André, Friesen, Zavagno, Clarke, Inoue, Chen, Lee, Soler, \& Kuffmeier}]{Pineda2023-di}
Pineda, J.~E., Arzoumanian, D., André, P., {et~al.} 2023

\bibitem[{{Poglitsch} {et~al.}(2010){Poglitsch}, {Waelkens}, {Geis}, {Feuchtgruber}, {Vandenbussche}, {Rodriguez}, {Krause}, {Renotte}, {van Hoof}, {Saraceno}, {Cepa}, {Kerschbaum}, {Agn{\`e}se}, {Ali}, {Altieri}, {Andreani}, {Augueres}, {Balog}, {Barl}, {Bauer}, {Belbachir}, {Benedettini}, {Billot}, {Boulade}, {Bischof}, {Blommaert}, {Callut}, {Cara}, {Cerulli}, {Cesarsky}, {Contursi}, {Creten}, {De Meester}, {Doublier}, {Doumayrou}, {Duband}, {Exter}, {Genzel}, {Gillis}, {Gr{\"o}zinger}, {Henning}, {Herreros}, {Huygen}, {Inguscio}, {Jakob}, {Jamar}, {Jean}, {de Jong}, {Katterloher}, {Kiss}, {Klaas}, {Lemke}, {Lutz}, {Madden}, {Marquet}, {Martignac}, {Mazy}, {Merken}, {Montfort}, {Morbidelli}, {M{\"u}ller}, {Nielbock}, {Okumura}, {Orfei}, {Ottensamer}, {Pezzuto}, {Popesso}, {Putzeys}, {Regibo}, {Reveret}, {Royer}, {Sauvage}, {Schreiber}, {Stegmaier}, {Schmitt}, {Schubert}, {Sturm}, {Thiel}, {Tofani}, {Vavrek}, {Wetzstein}, {Wieprecht}, \& {Wiezorrek}}]{Poglitsch2010}
{Poglitsch}, A., {Waelkens}, C., {Geis}, N., {et~al.} 2010, \aap, 518, L2

\bibitem[{Posch {et~al.}(2024)Posch, Alves, Mirét-Roig, Ratzenböck, Großschedl, Meingast, Swiggum, \& Konietzka}]{Posch2024-mz}
Posch, L., Alves, J., Mirét-Roig, N., {et~al.} 2024, A\&A in press

\bibitem[{Posch {et~al.}(2023)Posch, Miret-Roig, Alves, Ratzenböck, Großschedl, Meingast, Zucker, \& Burkert}]{Posch2023-in}
Posch, L., Miret-Roig, N., Alves, J., {et~al.} 2023, A\&A, 679, L10

\bibitem[{Preibisch \& Mamajek(2008)}]{Preibisch2008-ln}
Preibisch, T. \& Mamajek, E. 2008, Handbook of Star Forming Regions, 64, 235–370

\bibitem[{Ratzenbock {et~al.}(2023)Ratzenbock, Obermuller, Moller, Alves, \& Bomze}]{Ratzenbock2023-rw}
Ratzenbock, S., Obermuller, V., Moller, T., Alves, J., \& Bomze, I.~M. 2023, IEEE Trans. Vis. Comput. Graph., 29, 3855

\bibitem[{Ratzenböck {et~al.}(2023{\natexlab{a}})Ratzenböck, Großschedl, Alves, Miret-Roig, Bomze, \& {et al.}}]{Ratzenbock2023-sw}
Ratzenböck, S., Großschedl, J.~E., Alves, J., {et~al.} 2023{\natexlab{a}}, A\&A, 678, 71

\bibitem[{Ratzenböck {et~al.}(2023{\natexlab{b}})Ratzenböck, Großschedl, Möller, Alves, Bomze, \& Meingast}]{Ratzenbock2023-qb}
Ratzenböck, S., Großschedl, J.~E., Möller, T., {et~al.} 2023{\natexlab{b}}, A\&A, 677, A59

\bibitem[{Robitaille {et~al.}(2018)Robitaille, Scaife, Carretti, Haverkorn, Crocker, Kesteven, Poppi, \& Staveley-Smith}]{Robitaille2018-om}
Robitaille, J.-F., Scaife, A. M.~M., Carretti, E., {et~al.} 2018, A\&A, 617, A101

\bibitem[{Robitaille {et~al.}(2013)Robitaille, Tollerud, Greenfield, Droettboom, Bray, Aldcroft, Davis, Ginsburg, {Price-Whelan}, Kerzendorf, Conley, Crighton, Barbary, Muna, Ferguson, Grollier, Parikh, Nair, G{\"u}nther, Deil, Woillez, Conseil, Kramer, Turner, Singer, Fox, Weaver, Zabalza, Edwards, Bostroem, Burke, Casey, Crawford, Dencheva, Ely, Jenness, Labrie, Lim, Pierfederici, Pontzen, Ptak, Refsdal, Servillat, \& Streicher}]{2013A&A...558A..33A}
Robitaille, T.~P., Tollerud, E.~J., Greenfield, P., {et~al.} 2013, A\&A, 558, A33

\bibitem[{Rogers \& Pittard(2013)}]{Rogers2013-vx}
Rogers, H. \& Pittard, J.~M. 2013, MNRAS, 431, 1337

\bibitem[{Schlafly {et~al.}(2014)Schlafly, Green, Finkbeiner, Rix, Bell, Burgett, Chambers, Draper, Hodapp, Kaiser, Magnier, Martin, Metcalfe, Price, \& Tonry}]{Schlafly2014-tc}
Schlafly, E.~F., Green, G., Finkbeiner, D.~P., {et~al.} 2014, ApJ, 786, 29

\bibitem[{{Schneider} \& {Elmegreen}(1979)}]{Schneider1979}
{Schneider}, S. \& {Elmegreen}, B.~G. 1979, \apjs, 41, 87

\bibitem[{Shore(2007)}]{shoreAstrophysicalHydrodynamics2007}
Shore, S.~N. 2007, Astrophysical {{Hydrodynamics}} (Wiley)

\bibitem[{Skrutskie {et~al.}(2006)Skrutskie, Cutri, Stiening, Weinberg, Schneider, Carpenter, Beichman, Capps, Chester, Elias, Huchra, Liebert, Lonsdale, Monet, Price, Seitzer, Jarrett, Kirkpatrick, Gizis, Howard, Evans, Fowler, Fullmer, Hurt, Light, Kopan, Marsh, McCallon, Tam, Van~Dyk, \& Wheelock}]{Skrutskie2006-rn}
Skrutskie, M.~F., Cutri, R.~M., Stiening, R., {et~al.} 2006, AJ, 131, 1163

\bibitem[{{Smith}(2006)}]{Smith2006}
{Smith}, N. 2006, \mnras, 367, 763

\bibitem[{Smith {et~al.}(2020)Smith, Treß, Sormani, Glover, Klessen, Clark, Izquierdo, Duarte-Cabral, \& Zucker}]{Smith2020-xc}
Smith, R.~J., Treß, R.~G., Sormani, M.~C., {et~al.} 2020, MNRAS, 492, 1594

\bibitem[{Soler {et~al.}(2021)Soler, Beuther, Syed, Wang, Henning, Glover, Klessen, Sormani, Heyer, Smith, Urquhart, Yang, Su, \& Zhou}]{Soler2021-pu}
Soler, J.~D., Beuther, H., Syed, J., {et~al.} 2021, A\&A, 651, L4

\bibitem[{Soler {et~al.}(2018)Soler, Bracco, \& Pon}]{Soler2018-cc}
Soler, J.~D., Bracco, A., \& Pon, A. 2018, A\&A, 609, L3

\bibitem[{Soler {et~al.}(2022)Soler, Miville-Deschênes, Molinari, Klessen, Hennebelle, Testi, McClure-Griffiths, Beuther, Elia, Schisano, Traficante, Girichidis, Glover, Smith, Sormani, \& Treß}]{Soler2022-wh}
Soler, J.~D., Miville-Deschênes, M.-A., Molinari, S., {et~al.} 2022, A\&A, 662, A96

\bibitem[{{Struve} \& {Rudkjobing}(1948)}]{Struve1948}
{Struve}, O. \& {Rudkjobing}, M. 1948, \aj, 54, 51

\bibitem[{Tachihara {et~al.}(2000{\natexlab{a}})Tachihara, Abe, Onishi, Mizuno, \& Fukui}]{Tachihara2000-iu}
Tachihara, K., Abe, R., Onishi, T., Mizuno, A., \& Fukui, Y. 2000{\natexlab{a}}, PASJ, 52, 1147

\bibitem[{Tachihara {et~al.}(2000{\natexlab{b}})Tachihara, Mizuno, \& Fukui}]{Tachihara2000-gf}
Tachihara, K., Mizuno, A., \& Fukui, Y. 2000{\natexlab{b}}, ApJ, 528, 817

\bibitem[{Tachihara {et~al.}(2001)Tachihara, Toyoda, Onishi, Mizuno, Fukui, \& Neuhäuser}]{Tachihara2001-wr}
Tachihara, K., Toyoda, S., Onishi, T., {et~al.} 2001, PASJ, 53, 1081

\bibitem[{{Taylor}(2005)}]{2005ASPC..347...29T}
{Taylor}, M.~B. 2005, in ASPC, Vol. 347, Astronomical Data Analysis Software and Systems XIV, ed. P.~{Shopbell}, M.~{Britton}, \& R.~{Ebert}, 29

\bibitem[{Tritsis \& Tassis(2018)}]{Tritsis2018}
Tritsis, A. \& Tassis, K. 2018, Science, 360, 635

\bibitem[{{van Leeuwen}(2007)}]{van-Leeuwen2007}
{van Leeuwen}, F. 2007, \aap, 474, 653

\bibitem[{Vrba(1977)}]{Vrba1977-me}
Vrba, F. 1977, The Astronomical Journal, 82

\bibitem[{{Ward-Thompson} {et~al.}(1994){Ward-Thompson}, {Scott}, {Hills}, \& {Andre}}]{Ward-Thompson1994}
{Ward-Thompson}, D., {Scott}, P.~F., {Hills}, R.~E., \& {Andre}, P. 1994, \mnras, 268, 276

\bibitem[{Whitworth {et~al.}(2022)Whitworth, Priestley, \& Geen}]{Whitworth2022-zg}
Whitworth, A.~P., Priestley, F.~D., \& Geen, S.~T. 2022, MNRAS, 517, 4940

\bibitem[{{Wilking} {et~al.}(2008){Wilking}, {Gagn{\'e}}, \& {Allen}}]{Wilking2008}
{Wilking}, B.~A., {Gagn{\'e}}, M., \& {Allen}, L.~E. 2008, {Star Formation in the {$\rho$} Ophiuchi Molecular Cloud}, ed. B.~{Reipurth}, 351

\bibitem[{{Wood} {et~al.}(1992){Wood}, {Myers}, \& {Daugherty}}]{Wood1992}
{Wood}, D.~O.~S., {Myers}, P.~C., \& {Daugherty}, D.~A. 1992, in Bulletin of the American Astronomical Society, Vol.~24, American Astronomical Society Meeting Abstracts, 1200

\bibitem[{Zamora-Avilés {et~al.}(2019{\natexlab{a}})Zamora-Avilés, Ballesteros-Paredes, Hernández, Román-Zúñiga, Lora, \& Kounkel}]{Zamora-Aviles2019-ry}
Zamora-Avilés, M., Ballesteros-Paredes, J., Hernández, J., {et~al.} 2019{\natexlab{a}}, MNRAS, 488, 3406

\bibitem[{Zamora-Avilés {et~al.}(2019{\natexlab{b}})Zamora-Avilés, Vázquez-Semadeni, González, Franco, Shore, Hartmann, Ballesteros-Paredes, Banerjee, \& Körtgen}]{zamora-avilesStructureExpansionLaw2019}
Zamora-Avilés, M., Vázquez-Semadeni, E., González, R.~F., {et~al.} 2019{\natexlab{b}}, MNRAS, 487, 2200

\bibitem[{Zari {et~al.}(2016)Zari, Lombardi, Alves, Lada, \& Bouy}]{Zari2015}
Zari, E., Lombardi, M., Alves, J., Lada, C.~J., \& Bouy, H. 2016, A\&A, 587, A106

\bibitem[{Zucker {et~al.}(2017)Zucker, Battersby, \& Goodman}]{Zucker2017-nj}
Zucker, C., Battersby, C., \& Goodman, A. 2017, ApJ, 864, 153

\bibitem[{Zucker {et~al.}(2020)Zucker, Speagle, Schlafly, Green, Finkbeiner, Goodman, \& Alves}]{Zucker2020-gj}
Zucker, C., Speagle, J.~S., Schlafly, E.~F., {et~al.} 2020, A\&A, 633, A51

\bibitem[{Zucker {et~al.}(2019)Zucker, Speagle, Schlafly, Green, Finkbeiner, Goodman, \& Alves}]{Zucker2019-wr}
Zucker, C., Speagle, J.~S., Schlafly, E.~F., {et~al.} 2019, ApJ, 879, 125

\end{thebibliography}

\begin{appendix} %First online appendix
  % ----------------------------------------------------------------------
\section{Ionizing stars towards the Ophiuchus complex}
\label{sec:ioniz-stars}
In this section we summarize the properties of the massive stars in Upper-Sco, as they could be the main agents in shaping the local ISM via feedback forces. To quantify the energetics of the region we start by listing and estimating what fraction of the massive stars in Figure~\ref{fig:planck3color} are potentially interacting with the cloud complex. 
There are 20 massive ionizing stars (spectral type B3 and earlier) in the region, marked as star symbols in Figure~\ref{fig:planck3color}. Their main properties are listed in Table~\ref{tab:ionizing-stars}. The spectral type and V-band magnitude are taken from the Simbad astronomical database. We estimate for each star the hydrogen-ionizing photon luminosity (Q$_{\rm H}$) adopting the calibration in \cite{Smith2006}. 

For the distance we used the Hipparcos catalog distances \citep{van-Leeuwen2007} and equated parallax as distance, which for the fractional errors in the sample (all below 20\%) is a reasonable approximation \citep{Bailer-Jones2015}. Unfortunately, these 20 stars are too bright for a reliable parallax measurement in \textit{Gaia}.

Using the extended mid-infrared dust emission around these 20 massive stars as a proxy for the proximity between ionizing stars and the clouds, we searched for extended emission in both the WISE (bands 3 and 4) and the WHAM H$_\alpha$ surveys \citep{Haffner2003}. The results of this search are also listed in Table~\ref{tab:ionizing-stars}. All but the two faintest stars show WISE extended emission (open star symbols in Figure~\ref{fig:planck3color}) and most appear associated with extended H${\alpha}$ emission. A large fraction of these WISE nebula have crescent-shaped morphologies, or infrared arcs, resembling similar structures present in, e.g., $\sigma$ Ori or RCW 120 which indicate the presence of dust waves \citep{Ochsendorf2014} or stellar wind bubbles \citep{Mackey2015}. In summary, virtually all ionizing stars show some sort of WISE nebula at 12 $\mu$m, which implies proximity and interaction between the ionizing stars and the complex.

We estimate the present-day total ionizing photons luminosity in the region by summing the individual contributions in Table~\ref{tab:ionizing-stars} and obtain $\sim 10^{48.8}$ s$^{-1}$. This estimate is likely a lower limit due to unresolved binaries, which we correct for an ad hoc factor of two to estimate the present-day total ionizing photon luminosity in Upper-Sco to be Q$_{\rm H} \approx 10^{49}$ s$^{-1}$. 

\begin{table*}[h]
  \caption{Ionizing stars towards the Ophiuchus  complex (see Figure~\ref{fig:planck3color})}             % title of Table
  \label{tab:ionizing-stars}   
  % is used to refer this table in the text
  \centering                               % used for centering table
  \begin{tabular}{llccccc}        % centered columns (4 columns)
  \hline\hline                              % inserts double horizontal lines
  Name & Spectral Type & V & $\log Q_H/\SI{1}{s^{-1}}$ & Distance & WISE neb & H$\alpha$ neb\\
              &                          & (mag) & & (pc) &  &\\
  \hline
  Antares & M0.5Iab+B3V  & 0.91 & 46.37 & $170 \pm 29$ & Yes & Yes\\
  $\delta$ Sco & B0.2IVe      & 2.32 & 47.99 & $151 \pm 20$ & Yes & Yes\\
  $\zeta$ Oph & O9.2IV       & 2.56 & 48.36 & $112 \pm 2\phantom{0}$  & Yes & Yes\\
  $\beta$ Sco & B1V          & 2.62 & 47.28 & 124$\pm$12 & Yes & Yes\\
  $\tau$ Sco & B0.2V        & 2.81 & 47.70 & 145$\pm$12 & Yes & Yes\\
  $\sigma$ Sco & B1III+B1V    & 2.89 & 47.99 & 214$\pm$27 & Yes & Yes\\
  $\pi$ Sco & B1V+B2       & 2.91 & 47.40 & 180$\pm$20 & Yes & Yes\\
  $\rho$ Sco & B2IV-V       & 3.86 & 46.96 & 145$\pm$4  & Yes & No\\
  9 Sco & B1V            & 3.97 & 47.28 & 145$\pm$6  & Yes & No?\\
  $\nu$ Sco & B2V          & 4.00 & 46.80 & 145$\pm$17 & Yes & Yes\\
  $\chi$ Oph & B2Vne        & 4.43 & 46.80 & 161$\pm$6  & Yes & Yes\\
  13 Sco & B2V           & 4.57 & 46.80 & 147$\pm$4  & Yes & Yes\\
  2 Sco & B2.5Vn         & 4.59 & 46.59 & 154$\pm$12 & Yes & No?\\
  $\rho$ Oph & B2V-B3V      & 4.63 & 46.93 & 111$\pm$11 & Yes & Yes\\
  b Sco & B1.5Vn         & 4.64 & 47.05 & 152$\pm$6  & Yes & No?\\
  i Sco & B3V            & 4.79 & 46.37 & 127$\pm$4  & Yes & Yes?\\
  $\lambda$ Lib & B3V          & 5.03 & 46.37 & $\phantom{0} 95\pm 8\phantom{0}$& Yes & No\\
  V1040 Sco & B2V        & 5.40 & 46.80 & 131$\pm$6  & No & No\\
  HR5934 & B3V           & 5.85 & 46.37 & 141$\pm$6  & No & No\\
  %47 Lib & B2/B3V        & 5.96 & 46.58 & 236$\pm$48 & No & No\\
  HD147889 & B2III/IV    & 7.90 & 47.44 &  118$\pm$12& Yes& Yes\\
    \hline                                   %inserts single line
  \end{tabular}
  \tablefoot{Distances are taken from \cite{van-Leeuwen2007}. \cite{North2007} derived a closer distance of $174^{+23}_{-18}\,\mbox{pc}$ to $\sigma$ Sco from the orbital solution of the spectroscopic pair. Elias 2-9 (HD147889), in the vicinity of L1688, is extincted by about 5 mag of visual extinction \citep{Lombardi2008} and is then intrinsically brighter than the observed $\mathrm{V} = \SI{7.9}{mag}$.} 
  \end{table*}

\subsection{Key massive stars}
\label{sec:key-massive}

\begin{figure}
  \centering
  \includegraphics[width=1\hsize]{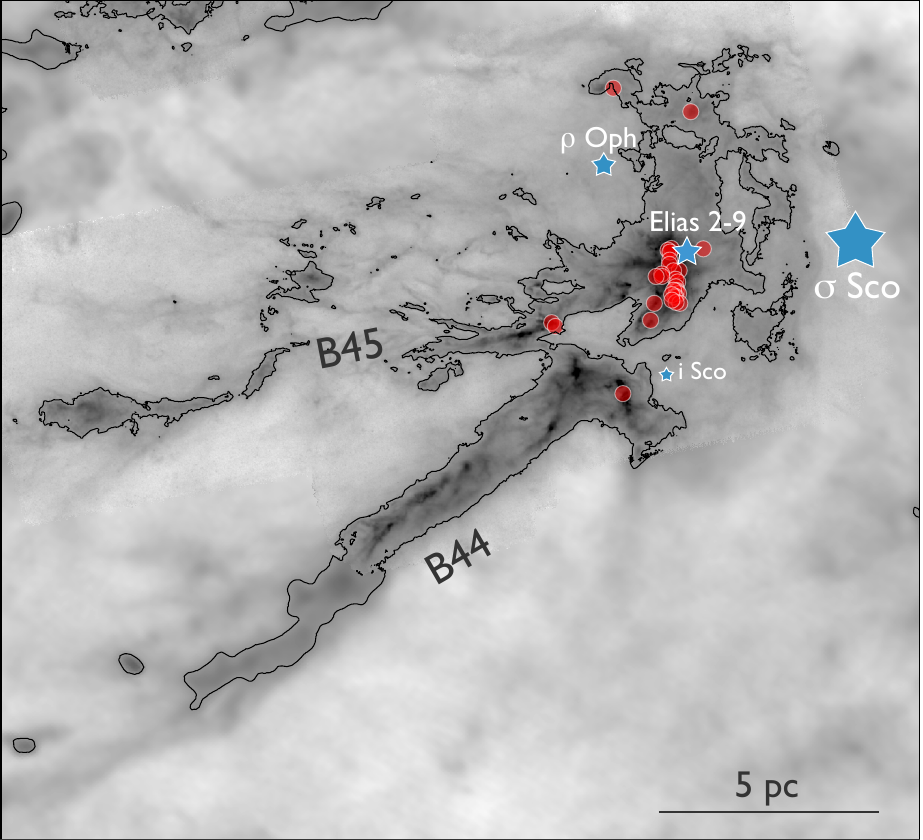}%
  \caption{The B44 and B45 filaments and the L1688 cloud containing a cluster of embedded protostars (red circles). The linear distribution of protostars in the L1688 cluster is orthogonal to the projected line-of-sight to $\sigma$-Sco, suggesting star formation in a compression front. The head of B45 (L1709) is radially aligned with $\sigma$ Sco, while B44 is radially aligned with Elias 2-9 (HD147889).}
\label{fig:OphClassic}
\end{figure}

Here we focus on the massive stars closest in projection and, when possible to tell, physically close to the main Ophiuchus clouds. In Figure~\ref{fig:OphClassic} we present a subset of Figure~\ref{fig:PHophmap_N} centered on the main Ophiuchus structures, that is, the L1688 clump and the filamentary structures B44 and B45. Blue stars represent massive ionizing stars in the region. The red circles represent the positions of deeply embedded Class I protostars and are taken from \cite{Kryukova2012}. We list below the closest ionizing stars to the Ophiuchus complex ordered by physical proximity.

\begin{enumerate}
    \item Elias 2-9 (or HD147889, a B2III star) powers a photodissociation region (PDR) \citep[e.g.,][]{Liseau2015-zp} on the western face of L1688 and is the facto closest massive star to L1688, at a projected distance of about 0.5 pc, measured from the angular distance between the star and the edge of the PDR. Elias 2-9 is similar in intrinsic brightness and mass to $\sigma$ Sco, but is seen through about 5 mag of visual extinction of L1688 material \citep{Lombardi2008-sa}, 

    \item $\rho$ Oph (B2V-B3V) is roughly at the same distance as the clouds \citep[e.g.,][]{Grasser2021-tt}, and is known for its associated reflection nebula that can be seen in the temperature map in Figure~\ref{fig:PHophmap_T}. The angular separation between this star and L1688 translates to about 1.5 pc.

    \item $\delta$ Sco (B0.2IVe, off image, see Figure~\ref{fig:planck3color}), the most massive B-star in Upper-Sco, is located at a distance of about 142 pc \citep{Ratzenbock2023-sw}, or at about the same distance as the Ophiuchus complex. This put it at about 13 pc from the Ophiuchus clouds.  

    \item $\sigma$ Sco (B1 III)   
    Hipparcos parallax suggests a distance of 214$^{+31}_{-25}$ pc \citep{van-Leeuwen2007}, while \cite{North2007-tt} propose a closer distance of 174$^{+23}_{-18}$pc based on the orbit of the binary pair. \cite{Ratzenbock2023-qb} estimate of the distance to the cluster likely to contain $\sigma$ Sco is 159$^{+7}_{-6}$ pc. This distance, derived from 544 \textit{Gaia} measurements, offers a reliable indicator of $\sigma$ Sco's true distance. Although one of the closest stars to the Ophiuchus clouds in projection, given its distance and angular separation, $\sigma$ Sco (B1 III) lies at about a distance of 19 pc from the clouds. 

\end{enumerate}

\newpage

\begin{table*}
\caption{Overview of the \textit{Herschel} data used in the paper}             % title of Table
\label{tab:A1}      % is used to refer this table in the text
\centering                          % used for centering table
\begin{tabular}{l c c c c c}        % centered columns (4 columns)
\hline\hline                 % inserts double horizontal lines
 Archival name          &  Obs. ID       &  {R.A.} &  {Dec.}
 &    Wavelengths ($\mu$m)    &  Obs. Date \\
\hline       %------------------------------------------------------------------------------inserts single line 
Sco-06             & 1342263838/9  &  245.39  & -19.86 &  250, 350, 500 & 2013-02-17 \\
Sco-05             & 1342263840/1  &  246.62  & -19.63 &  250, 350, 500 & 2013-02-17 \\
roph\_L1688        & 1342205093/4  &  246.70  & -24.18 &  250, 350, 500 & 2010-09-25 \\
Sco-04             & 1342263842/3  &  247.89  & -19.37 &  250, 350, 500 & 2013-02-18 \\
Sco-L43            & 1342263844/5  &  248.45  & -15.80 &  250, 350, 500 & 2013-02-18 \\
roph\_north\_stream& 1342214577/8  &  249.79  & -22.00 &  250, 350, 500 & 2011-02-20 \\
roph\_L1712        & 1342204088/9  &  250.03  & -24.12 &  250, 350, 500 & 2010-09-05 \\
Sco-01             & 1342267724/5  &  252.03  & -10.34 &  250, 350, 500 & 2013-03-16 \\
Sco-02             & 1342267726/7  &  252.04  & -12.29 &  250, 350, 500 & 2013-03-16 \\
Sco-03             & 1342267754/5  &  252.39  & -14.78 &  250, 350, 500 & 2013-03-17 \\
Sco-CB68           & 1342267756/7  &  253.87  & -15.84 &  250, 350, 500 & 2013-03-17 \\
\hline       %------------------------------------------------------------------------------inserts single line
\end{tabular}
\label{journalobs}
\end{table*}

\begin{figure}
  \centering
  \includegraphics[width=\hsize]{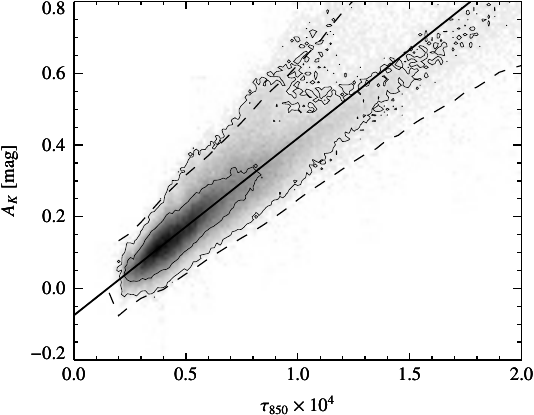}%
    \caption{Relationship between submillimeter optical-depth and NIR extinction in Ophiuchus. The best linear fit, used to calibrate the data, is shown together with the expected $3\sigma$ region, as calculated from direct error propagation in the extinction map.}
\label{fig:tau-nicest}
\end{figure}

\begin{figure}
  \centering
  \includegraphics[width=\hsize]{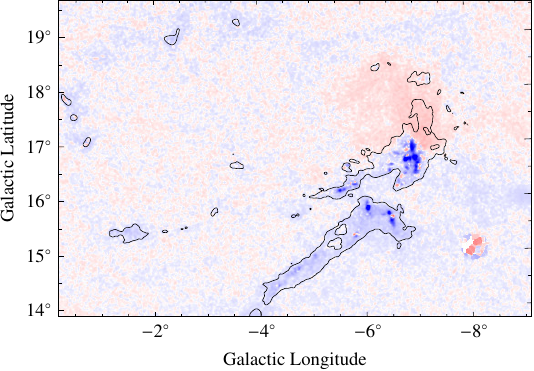}%
  \caption{Difference between the extinction predicted by \textit{Herschel} and the extinction measured with \textit{2MASS/Nicest}, with blue (red) indicating a positive (negative) difference.}
\label{fig:Herschel_nicest_map}
\end{figure}

\section{Error maps}

In this section, we show the error maps for the \textit{Herschel/Planck} column density and temperature maps presented in Figure~\ref{fig:PHophmap_N} and Figure~\ref{fig:PHophmap_T}. 

\begin{figure}
  \centering
  \includegraphics[width=\hsize]{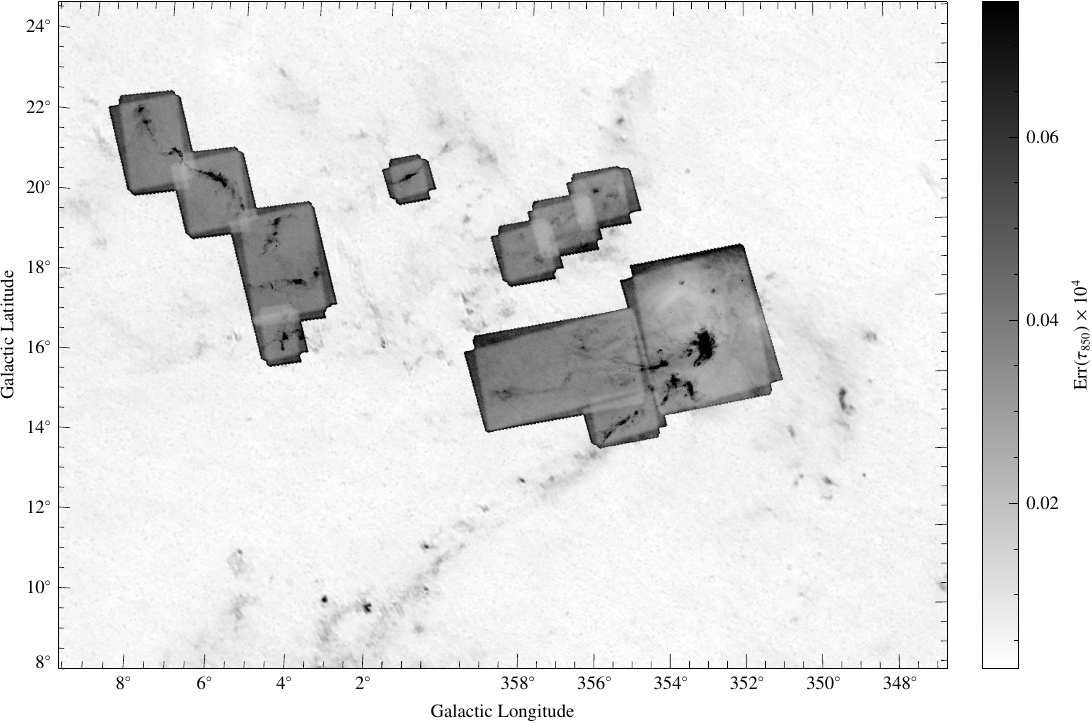}%
  \hspace{-\hsize}%
  \includegraphics[width=\hsize]{Figures/fig03a_labels5.pdf}%
  \caption{Error map for the optical-depth for the map reported in Figure~\ref{fig:PHophmap_N}. The figure shows the areas where the \textit{Herschel} data are available (higher errors). The resolution of the image varies from 5 arcmin (for the \textit{Planck} data, low error region) to 36 arcsec (for the \textit{Herschel}, higher error region). 
  }
  \label{fig:PHophmap_Nerr}
\end{figure}

\begin{figure}
  \centering
  \includegraphics[width=\hsize]{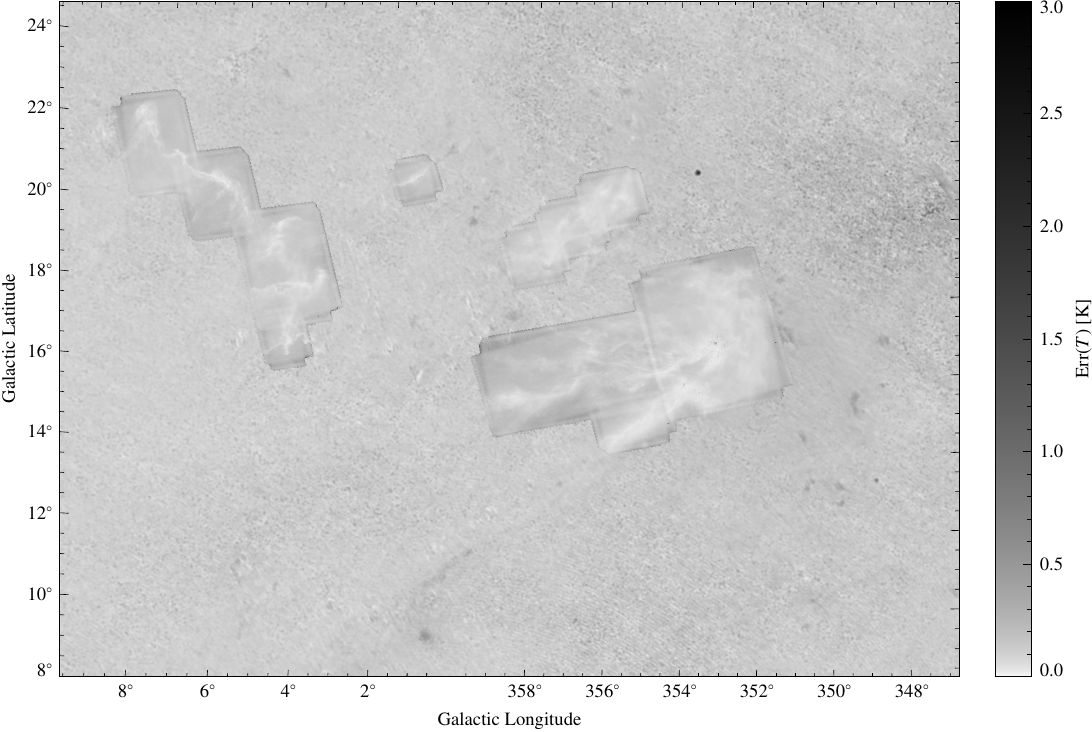}%
  \hspace{-\hsize}%
  \includegraphics[width=\hsize]{Figures/fig03a_labels5.pdf}%
  \caption{Error map for the effective dust-temperature map reported in Figure~\ref{fig:PHophmap_T}.}
  \label{fig:PHophmap_Terr}
\end{figure}

\section{SIMBAD Objects}
\object{Barn68}
\object{Ophiuchus Molecular Cloud}
\object{Lupus Cloud}
\object{Pipe Nebula}
\object{sigma Sco}
\object{pi Sco}
\object{delta Sco}
\object{zeta Oph}
\object{beta Sco}
\object{pi Sco}
\object{nu Sco}
\object{rho Sco}
\object{tau Sco}
\object{9 Sco}
\object{rho Oph}
\object{chi Oph}
\object{13 Sco}
\object{2 Sco}
\object{b Sco}
\object{i Sco}
\object{lambda Lib}
\object{V1040 Sco}
\object{HR5934}
\object{Antares}
\object{HD147889}
\object{barn 40}
\object{barn 44}
\object{barn 45}
\object{upper sco}
\object{L1712}

\end{appendix}
\end{document}